\documentclass[11 pt, oneside]{article}

\usepackage[pdftex]{graphicx}
\usepackage[toc,page]{appendix}

\usepackage{mathrsfs}
\usepackage{stmaryrd}
\usepackage{setspace}
\usepackage{colortbl}
\usepackage{bm}
\usepackage[normalem]{ulem}
\usepackage{subcaption}
\usepackage{graphicx}
\usepackage{multirow}
\usepackage{comment}

\usepackage[paperwidth=8.5in, paperheight=11in, top=1in, bottom=1in, left=1in, right=1in]{geometry}

\usepackage{endnotes}
\let\footnote=\endnote

\usepackage{geometry}
\usepackage{graphicx}
\usepackage{setspace}
\usepackage{natbib}
\usepackage[scaled=0.9]{helvet}

\usepackage{booktabs}

\usepackage{amsmath,amssymb,amsthm,mathrsfs,amsfonts,dsfont}

\usepackage{multirow}
\usepackage{algorithm}
\usepackage{algpseudocode} 
\usepackage{enumerate}
\usepackage[affil-sl,auth-sc]{authblk}
\usepackage{color}

\newcommand{\bY}{\mathbf{Y}}
\newcommand{\bW}{\mathbf{W}}
\newcommand{\bA}{\mathbf{A}}
\newcommand{\bI}{\mathbf{I}}

\newcommand{\bF}{\mathbf{F}}
\newcommand{\bPhi}{\mathbf{\Phi}}
\newcommand{\bphi}{\boldsymbol{\phi}}
\newcommand{\bV}{\mathbf{V}}

\newcommand{\bv}{\mathbf{v}}
\newcommand{\valpha}{\boldsymbol{\alpha}}
\newcommand{\vbeta}{\boldsymbol{\beta}}

\usepackage{natbib}
 \bibpunct{(}{)}{;}{a}{,}{,}
 \def\newblock{\ }%

\usepackage{color}

\newtheorem{theorem}{Theorem}
\theoremstyle{definition} 
\theoremstyle{remark} 
\theoremstyle{remark} 

\date{}

\begin{document}

\title{A Reproducing Kernel Hilbert Space Approach to Functional Calibration of Computer Models}
\date{}
\author{Rui Tuo, Shiyuan He, Arash Pourhabib, Yu Ding and Jianhua Z. Huang}
\date{}

\maketitle
\pagenumbering{roman}

\begin{abstract}
This paper develops a frequentist solution to the functional calibration problem, where the value of a calibration parameter in a computer model is allowed to vary with the value of control variables in the physical system. The need of functional calibration is motivated by engineering applications where using a constant calibration parameter results in a significant mismatch between outputs from the computer model and the physical experiment. Reproducing kernel Hilbert spaces (RKHS) are used to model the optimal calibration function, defined as the functional relationship between the calibration parameter and control variables that gives the best prediction. This optimal calibration function is estimated through  penalized least squares with an RKHS-norm penalty and using physical data. An uncertainty quantification procedure is also developed for such estimates.
Theoretical guarantees of the proposed method are provided in terms of prediction consistency and consitency of estimating the optimal calibration function. The proposed method is tested using both real and synthetic data and exhibits more robust performance in prediction and uncertainty quantification than the existing parametric functional calibration method and a state-of-art Bayesian method.


\noindent \textbf{Short title:} Functional calibration

\noindent Computer code and data are available at GitHub, http://github.com/shiyuanhe/calibration
\end{abstract}

\newpage
\begin{center}
\textbf{Author's Footnote:}
\end{center}

Shiyuan He (Email: \texttt{heshiyuan@ruc.edu.cn}) is Assistant Professor, Institute of Statistics and Big Data, Renmin University of China, 59 Zhongguancun Rd., Haidian District, Beijing 100872, China,
Arash Pourhabib (Email: \texttt{arash.pourhabib@okstate.edu}) is Adjunct Assistant Professor, Oklahoma State University, Stillwater, OK 74078,
Rui Tuo (Email: \texttt{ruituo@tamu.edu}) is Assistant Professor, Yu Ding (Email: \texttt{yuding@tamu.edu}) is Professor, Department of Industrial and Systems Engineering,
Jianhua Z. Huang (Email: \texttt{jianhua@stat.tamu.edu}) is Professor, Department of Statistics, Texas A\&M University, College Station, TX 77843. He's work was supported by NSFC Project 11801561.
Ding's work was partially supported by NSF grants
CMMI-1545038, IIS-1849085, CCF-1934904. Huang's work was partially supported by NSF grants DMS-1208952, IIS-1900990, CCF-1956219. The first two authors, Tuo and He, made equal contributions to the paper. Corresponding author: Jianhua Huang.
\newpage

\pagenumbering{arabic}
\setcounter{page}{1}

\section {Introduction }\label{sec_intro}

To understand a physical system, one can conduct physical experiments by feeding a set of inputs to the system and observing the output. These inputs are called \emph{control variables} to the system. The hope is that by learning the input-output relation of the system, in the future one can predict the output for any set of inputs that may be fed into the system. Because conducting physical experiments is usually very costly and inconvenient, computer simulation or using computer models is a common practice~\citep{santner2003design}. A computer model attempts to use a set of mathematical formulas to mimic the input-output relation of the physical system and can be implemented through computer codes. There are usually a set of parameters in the computer model that represent intrinsic properties of the physical system. Different from control variables that one can determine and measure before an experiment, the computer model parameters are unobservable or unmeasurable. To determine the value of these parameters, one first obtains some data in the form of input-output pairs from physical experiments, and then adjusts the values of computer model parameters so that the computer model generates similar outputs for a given set of inputs in the physical experiments. This process is called \textit{calibration}, and we refer to the computer model parameters as \emph{calibration parameters}. \cite{kennedy2001bayesian} is a renowned work on statistical framework for calibration of computer models, which also contains a review of earlier work on the subject. Related references in this field include  \cite{higdon2004combining,higdon2008computer,higdon2013computer,bayarri2007framework,Bayarri2007,tuo2014efficient,Tuo2014calibration,joseph2009statistical,joseph2015engineering,Wong2017,plumlee2017bayesian,gu2018scaled, plumlee2019computer,tuo2019adjustments,xie2020bayesian,wang2020effective} and the references therein.

Commonly used approaches to calibration are based on the assumption that there is one constant value for the calibration parameter, and the goal of calibration is to estimate that value, or find the closest possible point~\citep{han2009simultaneous,Tuo2014calibration}. However, in many complex physical systems, there may exist a functional relationship between calibration parameters and control variables. 
For example, in the resistance spot welding process discussed in~\citet{bayarri2007computer}, or in the poly-vinyl alcohol (PVA)-treated buckypaper fabrication process~\citep{Pourhabib2014bp}, engineering knowledge suggests that calibration parameters, i.e., contact resistance in the former and the PVA absorption rate in the latter, are not fixed but a function of control variables. In these examples, any attempts that seek to calibrate the computer model by finding the ``best'' constant value for the calibration parameter will result in a significant mismatch between the outputs from the computer model and the physical experiment. Therefore, appropriate calibration of the computer model in such cases should allow the values of the calibration parameters to vary with the value of the input variables. To highlight the functional relationship, we refer to the calibration parameters as the \emph{functional calibration parameters} or \textit{calibration functions}. If the calibration function has a known form with an unknown Euclidean parameter, one can develop a parametric solution to the functional calibration problem~\citep{bayarri2007computer, Pourhabib2014bp, Atamturktur2015}, and the theoretical framework of \cite{tuo2014efficient,Tuo2014calibration} can be employed to find the optimal calibration parameter. However, such parametric solutions have obvious limitations, since the form of the calibration function is usually unavailable. 

The goal of this paper is to provide a theoretical framework for a nonparametric solution to the functional calibration problem and to develop corresponding computational methods and uncertainty quantification procedures. We formulate the problem such that it allows the data to determine the functional relationship between the calibration parameter and control variables, instead of imposing a parametric form for the relationship \textit{a priori}. We utilize reproducing kernel Hilbert spaces to model the functional relationship and employ the penalized least squares for estimation. We devise a Gauss-Newton type of algorithm for computation and establish frequentist properties of our estimation procedure. We also develop an uncertainty quantification procedure by borrowing ideas from the smoothing splines literature. Our framework treats the computer model as an approximation to the physical system and do not assume there exists a ``true'' underlying calibration function. The optimal calibration function is defined operationally through an optimization problem to give the best prediction.
Interestingly, when there are multiple functional calibration parameters, there is a possibility that the optimal calibration functions are not uniquely defined (see the discussion in Section~\ref{sec:local-calibration}), but our procedure still yields consistent prediction (Section~\ref{sec:pred-consistency}) and our numerical results show good performance of our method in prediction (Section~\ref{sec_res}).

Upon completion of an earlier draft of this work, we became aware of two related papers that tackled the same problem using the Bayesian framework. \citet{Plumlee2016} provided a Bayesian solution to the functional calibration problem in a specific application. \cite{brown2018nonparametric} developed a general Bayesian framework with a software implementation. Both papers utilized Gaussian Process (GP) priors on the unknown calibration functions and applied the Markov Chain Monte Carlo simulation for computation of the posterior distribution. While theoretical justification is lacking for these existing Bayesian functional calibration methods, we are able to establish an asymptotic theory of consistency and rates of convergence to provide some theoretical guarantees to our frequentist method. Our method still provides good prediction and uncertainty quantification when the optimal functional calibration parameter is not uniquely defined (see Tables~\ref{tbl:simu3} and~\ref{tbl:simu4}). The existing Bayesian methods have not considered this challenging situation. This paper also contains a comparative simulation study that is more comprehensive than those in the published works. Empirical evidence shows that the proposed frequentist approach tends to outperform existing Bayesian methods on the examples considered.

The rest of this paper is organized as follows. Section \ref{sec_cal} provides basic concepts regarding calibration, the formulation of the functional calibration problem, and a solution procedure through penalized least squares. Section~\ref{sec_the} provides theoretical properties of the proposed method. Section~\ref{sec:computation} presents a computational algorithm for solving the penalized least squares problem and derives the GCV criterion for penalty parameter selection. Section~\ref{sec:UQ} develops an uncertainty quantification procedure for estimation of calibration function and for prediction. Section~\ref{sec_res} presents the results for a simulation study. We illustrate the proposed method on real data in Section~\ref{sec_buckypaper}.

{\rm Notations.} For a vector $\bm a$, let $|\bm a|$ denote its Euclidean norm. For a function $f(\bm x)$ defined on a domain $\mathcal{X}$, let $\|f\|_{L_2(\mathcal{X})} = \{\int_{\mathcal{X}} f^2(\bm x)\, d\bm x\}^{1/2}$ denote its $L_2$ norm. For a vector of functions $\bm f = (f_1, \dots, f_r)^T$, denote $\|\bm f\|_{L_2} = \{\sum_{i=1}^r \|f_i\|^2_{L_2(\mathcal{X})} \}^{1/2}$.


\section{Functional Calibration: Problem Formulation}\label{sec_cal}

Consider a physical system that gives a vector of deterministic responses $\bm \zeta(\bm x) \in \mathbf{R}^r $ when there is a vector of inputs (called control variables), $\bm x\in\mathcal{X}$, where $\mathcal{X}$ is a convex and compact subset of $\mathbf{R}^d$. To learn the response function $\bm\zeta(\cdot)$, we conduct physical experiments at design points $\mathcal{D}=\{\bm x_1,\ldots,\bm x_n\}\subset\mathcal{X}$ and observe the corresponding response vectors, denoted as $\bm y_1^p,\ldots, \bm y_n^p$, where superscript $p$ stands for ``physical.'' A functional relationship exists between the input $\bm x$ and the response $\bm \zeta(\bm x)$, but due to measurement noises we only observe a noisy version of the response. We assume that
\begin{eqnarray}\label{eq:data-model}
\bm y_i^p = \bm \zeta(\bm x_i)+\bm e_i, \qquad i=1, \dots, n,
\end{eqnarray}
where $\bm e_i$'s are i.i.d zero-mean random vectors. We call each pair $(\bm x_i, \bm y_i^p)$ a data point and the set $\{(\bm x_i, \bm y_i^p), i=1, \dots, n\}$ the physical dataset.

\subsection{Optimal constant calibration}

Exploring the response function through mere physical experimentation can be extremely costly and time consuming. As such, a computer model is often utilized to simulate the physical system.  The computer model takes $(\bm x,\bm \theta)$ as the input and yields the computer model response $\bm y^s(\bm x,\bm \theta)$, where $\bm \theta\in\Theta$ denotes a calibration parameter, $\Theta$ is a subset of $\mathbf{R}^q$, and superscript $s$ stands for ``simulated,'' referring to simulated from the computer model. The computer model response $\bm y^s(\bm x,\bm \theta)$ is computable by running computer codes when a calibration parameter ${\bm \theta}$ and an input $\bm x$ are given.
An appealing feature of using computer models is that one can run the computer codes for the computer model for any feasible  combination of control variables and calibration parameters, at much lower expense than running physical experiments. A fundamental problem for using computer models is \emph{calibration}, i.e., the problem of finding a suitable value for the calibration parameters so that the computer outputs match well those from the physical experiment for the same inputs or values of control variables.

Being built based on simplifying assumptions about the physical system, a computer model may not perfectly match the physical system. Therefore, we should not expect there exists a value $\bm \theta$ such that the computer model response $\bm y^s(\bm x,\bm \theta)$ exactly equals 
the physical system response $\bm \zeta(\bm x)$. The objective in calibration is to adjust the imperfect computer model, through changing the values of the calibration parameter, so that it adequately represents the physical system. Following \cite{tuo2014efficient,Tuo2014calibration} and \cite{Wong2017}, we define the optimal value $\bm \theta^*$ of the calibration parameter $\bm \theta$ to be the value that minimizes the $L_2$ distance between $\bm \zeta(\bm x)$ and $\bm y^s(\bm \cdot,\bm\theta)$, i.e.,
\begin{eqnarray}\label{least distance}
\bm \theta^*=\operatorname*{argmin}\limits_{\bm \theta\in\Theta}\|\bm \zeta(\bm x)- \bm y^s(\bm x,\bm \theta)\|^2_{L_2(\mathcal{X})}
= \operatorname*{argmin}\limits_{\bm \theta\in\Theta}\int_\mathcal{X} |\bm \zeta(\bm x)-\bm y^s(\bm x,\bm \theta)|^2 \, d\bm x.
\end{eqnarray}
Empirically, the optimal calibration parameter can be estimated by 
solving a similar optimization problem as \eqref{least distance} where the integral for the $L_2$ distance is replaced by sample average for a given physical dataset, and each ${\bm \zeta}({\bm x}_i)$ by its noisy observation ${\bm y}_i^p$.
This empirical calibration method is called the ordinary least squares method in Section 4 of \citet{tuo2014efficient}.

\subsection{Optimal functional calibration}\label{sec:local-calibration}
In the above discussion, we look for one single value of the calibration parameter such that the computer model best represents the physical experiment. We now consider the situation that there is a functional relationship between the calibration parameters and control variables. Our goal is to learn this functional relationship using physical data. 
Extending \eqref{least distance}, we define the \emph{optimal calibration function} for the functional calibration problem as
\begin{equation}\label{eq:mindis}
\begin{split}
\bm\theta^*(\cdot)=\operatorname*{argmin}_{\bm\theta(\cdot)} \int_{\mathcal{X}}|\bm \zeta(\bm x)- \bm y^s(\bm x,\bm \theta(\bm x))|^2\,d\bm x,\\
\text{subject to: } \bm \theta(\bm x)\in\Theta \text{ for all } \bm x\in\mathcal{X}.
\end{split}
\end{equation}
We refer to $\bm \theta^*(\bm x)$ as the \emph{optimal functional calibration parameter} and $\bm \zeta^*(\bm x) = \bm y^s(\bm x,\bm \theta^*(\bm x))$ as the \emph{optimal prediction} at location $\bm x$. Similar to \eqref{least distance}, here we do not assume a true value/form for the calibration function, but we seek a function that minimizes the discrepancy between the outputs of computer model and physical experiments.
In fact, $\bm \theta^*(\cdot)$ can be defined pointwise: $\bm \theta^*(\bm x)$ is the minimizer of the function
$h_{\bm x}(\bm \theta) = |\bm \zeta(\bm x)-\bm y^s(\bm x,\bm \theta)|^2$ for each $\bm x$.
Since $\Theta$ is compact, by assuming that $h_{\bm x}(\bm\theta)$ is a continuous function, the minimizer  $\bm \theta^*(\bm x)$ always exists. 

The optimal functional calibration parameter $\bm \theta^*(\bm x)$ may not be unique, although the optimal prediction $\bm \zeta^*(\bm x)$ is uniquely defined. Section~\ref{sec_res} presents some examples when there are multiple or even continuum many minimizers $\bm \theta^*(\bm x)$ for \eqref{eq:mindis}; see simulation settings 3 and 4. A key condition for the uniqueness of $\bm \theta^*(\bm x)$ is $r \geq q$, i.e., the number of response variables is no fewer than the number of functional calibration parameters. To see this, note that $\bm \theta = \bm \theta^*(\bm x)$ solves the system of $r$ equations $\bm \zeta^*(\bm x) = \bm y^s(\bm x, \bm \theta)$, and thus in order for the system to have a unique solution, the number of equations should be no fewer than the number of parameters. Under mild conditions, the existance and uniqueness of the solution is ensured by the inverse function theorem or the implicit function theorem.
On the other hand, since the primary goal of computer experiment is usually on prediction of outcomes of physical experiments, the uniqueness of $\bm \theta^*(\bm x)$ is not that a big concern.

\subsection{Empirical functional calibration}

Recall that what we observe are noise-contaminated physical responses $\bm y^p_i$ at a finite number of values of control variables $\bm x_i\in\mathcal{D}$. We need to learn the optimal calibration function using the physical data $\{\bm x_i, \bm y^p_i\}$. One could replace the integral in \eqref{eq:mindis} by a summation over the design points $\bm x_i$'s, replace each ${\bm \zeta}({\bm x}_i)$ by its noisy observation ${\bm y}_i^p$,
and solve the corresponding minimization problem. However, this minimization problem does not have a unique solution since one can vary the values of the function at points other than the design points and do not change the value of the minimizing objective function. Therefore, to make $\bm \theta^*(\cdot)$ estimable, we have to postulate certain assumptions on $\bm\theta^*(\cdot)$. A common assumption is to suppose $\bm\theta^*$ has certain degree of smoothness. Therefore we shall restrict our attention to elements in a suitably chosen space of smooth functions. Specifically, we consider \textit{native spaces}, which is a generalization of \textit{reproducing kernel Hilbert spaces} (RKHS). We refer to \cite{wendland2005scattered} for the necessary mathematical background of RKHS and native spaces.

Let $\Phi$ be a conditionally positive definite function over $\mathcal{X}\times\mathcal{X}$ and $\mathcal{N}_\Phi(\mathcal{X})$ be the native space generated by $\Phi$ with its native semi-norm $\|\cdot\|_{\mathcal{N}_{\Phi}}$. In general, the native semi-norm of a function $f$, $\|f\|_{\mathcal{N}_{\Phi}}$ is a measure of roughness of $f$, with a larger value indicating a rougher function. We define our empirical calibration function $\hat{\bm\theta}(\cdot)$ to be
\begin{equation}
\begin{split}\label{eq:LC_gen}
\hat{\bm\theta}(\cdot):=\operatorname*{argmin}\limits_{\bm\theta(\cdot) = (\theta_1(\cdot), 
\dots, \theta_q(\cdot))^T } \frac{1}{n}\sum_{i=1}^n \left|\bm y_i^p-\bm y^s(\bm x_i,\bm\theta(\bm x_i))\right|^2+\lambda\sum_{j=1}^q\|\theta_j\|_{\mathcal{N}_{\Phi}}^2,\\
\text{subject to: } \bm \theta(\bm x)\in\Theta \text{ for all } \bm x\in\mathcal{X},
\end{split}
\end{equation}
where the first term of the objective function measures the goodness-of-fit of the computer model's responses to the physical responses, the semi-norm $\|\theta_j\|_{\mathcal{N}_{\Phi}}$ measures the roughness of $\theta_j$, and
$\lambda>0$ is a regularization/penalty parameter that balances the data fit with function smoothness. It is necessary that the solution $\hat{\bm\theta}(\cdot) = (\hat\theta_1(\cdot), \dots, \hat\theta_q(\cdot))$ must satisfy $\|\hat{\theta}_j\|_{\mathcal{N}_{\Phi}}<\infty$ so that its components are in the native space generated by $\Phi$. We call our proposed approach the nonparametric functional calibration, since we do not assume a parametric form for the calibration function $\bm\theta(\cdot)$. The prediction of the outcome at a new location $\bm x$ is obtained by the plug-in estimator $\bm y^s(\bm x, \hat{\bm \theta}(\bm x))$.

So far we have assumed the computer model is cheap, i.e., $\bm y^s(\bm x,\bm \theta)$ can be evaluated at any arbitrary point $(\bm x,\bm \theta)\in\mathcal{X}\times\Theta$ with almost no cost. However, in practice there is always computational cost associated with running a computer code. When the computational cost cannot be ignored when comparing with the physical experiments, we say that the computer model is expensive. To deal with the case of an expensive computer model,
we can evaluate the computer model only at a set of computer design points $\mathcal{G}=\{(\bm x_1,\bm \theta_1),\ldots,(\bm x_m,\bm \theta_m)\}$ and then build an emulator, or a surrogate model, based on the evaluation of the computer model at these design points~\citep{santner2003design}. Denote such an emulator as $\hat{\bm y}_m^s(\bm x,\bm \theta)$ which is obtained using $\{(\bm x_j, \bm \theta_j, \bm y^s(\bm x_j,\bm \theta_j)), j=1, \dots, m\}$. Simply replacing the computer model $\bm y^s(\bm x,\bm \theta)$ in \eqref{eq:LC_gen} by the emulator, we obtain
\begin{equation}
\begin{split}\label{eq:LC_gen_exp}
\hat{\bm \theta}_{(m)}(\cdot):=\operatorname*{argmin}\limits_{\bm\theta(\cdot) = (\theta_1(\cdot), 
\dots, \theta_q(\cdot))^T } \frac{1}{n}\sum_{i=1}^n \left|\bm y^p_i-\hat{\bm y}^s_m(\bm x_i,\bm \theta(\bm x_i))\right|^2+\lambda\sum_{j=1}^q\|\theta_j\|_{\mathcal{N}_{\Phi}}^2,\\
\text{subject to: } \bm \theta(\bm x)\in\Theta \text{ for all } \bm x\in\mathcal{X}.
\end{split}
\end{equation}
We could select the design points $\mathcal{G}$ such that $m=n$, and for any $\bm x_i\in\mathcal{D}$ there exists $\bm \theta_i$ such that $(\bm x_i,\bm \theta_i)\in\mathcal{G}$, i.e., we build the emulator based on the same design points augmented by the $\bm\theta$'s. However, in most applications, even expensive computer codes are much cheaper than their associated physical experiments and thus it is often reasonable to choose $m\gg n$.

{\bf Remark.} The criterion functions in the optimization problems \eqref{eq:LC_gen} and \eqref{eq:LC_gen_exp} can be compared with the logarithm of the posterior density presented on pages 727--728 of~\cite{brown2018nonparametric}. While the two terms in \eqref{eq:LC_gen} and \eqref{eq:LC_gen_exp} also appeared in the Bayesian formulation, the log posterior density has a few extra terms that are the result of prior specification.


\section{Theoretical Properties}\label{sec_the}
In this section, we develop some asymptotic properties for the proposed nonparametric functional calibration method. Here ``asymptotic'' means that the sample size of the physical observations tends to infinity and the error of the emulator tends to zero. 
In Section \ref{sec:pred-consistency}, we study the prediction consistency when the optimal calibration function is not required to be unique. In Section \ref{sec_estimation}, we show the consistency of estimating the optimal calibration function when it is uniquely defined. The rates of convergence are also discussed for both scenarios.

For the rest of this section, we fix a conditionally positive definite kernel function $\Phi$ defined  on $\mathcal{X}\times \mathcal{X}$. Let $\mathcal{N}(\mathcal{X})$ be the \textit{native space} generated by the kernel $\Phi$ with its native semi-norm $\|\cdot\|_{\mathcal{N}}$ (we drop the subscript
$\Phi$ in $\mathcal{N}_{\Phi}$ to simplify the notation).
Recall that $\Theta$, the domain of $\theta$, is a subset of $\mathbf{R}^q$.
Denote $\mathcal{N}^{\Theta}:=\{\bm f=(f_1,\ldots,f_q):f_j\in\mathcal{N}(\mathcal{X}),j=1,\ldots,q,\bm f(x)\in\Theta \text{ for all } x\in\mathcal{X}\}$. The functional calibration problems \eqref{eq:LC_gen} or \eqref{eq:LC_gen_exp} can be cast into a unified form as
\begin{eqnarray}
\hat{\bm \theta}_n:=\operatorname*{argmin}\limits_{\bm \theta\in\mathcal{N}^{\Theta}}\frac{1}{n}\sum_{i=1}^n \left\|\bm y_i^p-\hat{\bm y}_n^s(\bm x_i,\bm \theta(\bm x_i))\right\|^2+\lambda_n\sum_{j=1}^q \|\theta_j\|^2_{\mathcal{N}},\label{LC}
\end{eqnarray}
for a sequence of smoothing parameters $\lambda_n>0$, where $\{\hat{\bm y}^s_n\}$
is a sequence of emulators for $\bm y^s$ with increasing accuracies as $n\to \infty$. For the
case of cheap codes, letting $\hat{\bm y}_n^s(\bm x_i,\bm \theta(\bm x_i)) = {\bm y}^s(\bm x_i, \bm \theta(\bm x_i))$ in
\eqref{LC} gives \eqref{eq:LC_gen}; for the case of expensive codes, letting
$\hat{\bm y}_n^s(\bm x_i,\bm \theta(x_i)) = \hat{\bm y}^s_m(\bm x_i,\bm \theta(\bm x_i))$ in \eqref{LC} gives \eqref{eq:LC_gen_exp}.
In the latter case, we allow $m$ to depend on $n$. It is worth noting that if the solution of (\ref{LC}) is not unique and we focus on prediction, our theory works for an arbitrary choice from possible solutions.

\subsection{Prediction consistency}\label{sec:pred-consistency}
According to \eqref{eq:mindis}, the computer model response $\bm y^s(\bm x, \bm \theta^*(x))$, equipped with the optimal calibration function $\bm \theta^*(\cdot)$, provides the best possible prediction of the physical system response $\bm \zeta(\bm x)$. The corresponding prediction error measured in $L_2$ distance is 
\[
\mathrm{PE}(\bm \theta^*) = \int_{\mathcal{X}}|\bm \zeta(\bm x)- \bm y^s(\bm x,\bm \theta^*(\bm x))|^2\,d\bm x.
\]
The $L_2$ norm of prediction error corresponding to the empirical calibration function defined in \eqref{LC} is
\[
\mathrm{PE}(\hat{\bm \theta}_n)= \int_{\mathcal{X}}|\bm \zeta(\bm x)- \hat{\bm y}^s_n(\bm x,\hat{\bm \theta}(\bm x))|^2\,d\bm x.
\]
The difference between these two quantities tells us how well the emprical calibration function performs in terms of prediction.
This subsection establishes a prediction consistency result, which states that $\mathrm{PE}(\hat{\bm \theta}_n)- \mathrm{PE}(\bm \theta^*) \to 0$  when the sample size $n \to \infty$.  
Our result does not require the uniqueness of $\bm \theta^*$.  Note that the value of $\mathrm{PE}(\bm \theta^*)$ is always uniquely defined and does not depend on the choice of feasible $\bm \theta^*$.

Before stating our results, we introduce some technical conditions.
Call a $r$-dimensional random vector $\bm \xi$ sub-Gaussian, if there exists $\varsigma>0$ such that
\begin{eqnarray}
E[\exp \{\bm \alpha^T(\bm \xi-\mathbb{E}\bm \xi)\}]\leq \exp(\varsigma^2\|\bm \alpha\|^2/2),\label{subGaussian}
\end{eqnarray}
for all $\bm \alpha\in\mathbf{R}^{r}$. Here the constant $\varsigma$ is referred to as the sub-Gaussian parameter.

{\bf Condition 1.} The design points $\bm x_i$'s are randomly drawn from the uniform distribution over $\mathcal{X}$, and the noise vectors $\bm e_i$'s are independent from a sub-Gaussian distribution with mean zero and sub-Gaussian parameter $\sigma$, for $i=1,\ldots,n$.
Moreover, $\{\bm x_i\}_{i=1}^n$ and $\{\bm e_i\}_{i=1}^n$ are mutually independent.

{\bf Condition 2.} The computer model is a smooth function of its inputs in the sense that
$\bm y^s\in C^2(\mathcal{X}\times\Theta)$, the space of twice continuously differentiable functions defined on $\mathcal{X}\times\Theta$.

Define the covering number $N(\delta,\mathcal{S},d)$ of the set $\mathcal{S}$ as the smallest value of $N$ for which there exist functions $f_1,\ldots,f_N$, such that for each $f\in\mathcal{S}$, $d(f,f_j)\leq\delta$ for some $j\in\{1,\ldots,N\}$. Let $\mathcal{N}(\mathcal{X},\rho):=\{f\in\mathcal{N}(\mathcal{X}):\|f\|_\mathcal{N}\leq \rho\}$ for $\rho>0$. 

{\bf Condition 3.} The covering number of $\mathcal{N}(\mathcal{X},\rho)$ satisfies
\begin{eqnarray}
\log N(\epsilon,\mathcal{N}(\mathcal{X},\rho),L_\infty(\mathcal{X}))\leq \left(\frac{C \rho}{\epsilon}\right)^{\frac{d}{\nu}},\label{entropy}
\end{eqnarray}
for some $\nu> d/2$ and a constant $C$ independent of $\epsilon$ and $\rho$.

Condition~3 gives a constraint on the size of the reproducing kernel Hilbert space. Normally the value of $\nu$ depends highly on the smoothness of the kernel function. Consider the Mat\'{e}rn family of kernel functions \citep{stein1999interpolation},
\begin{eqnarray}\label{matern}
\Phi(s,t)=\Phi_{\upsilon,\phi}(s,t)=\frac{1}{\Gamma(\upsilon)2^{\upsilon-1}}\left(2\sqrt{\upsilon}\phi \|s-t\|\right)^\upsilon K_\upsilon\left(2\sqrt{\upsilon}\phi\|s-t\|\right),
\end{eqnarray}
where $K_\upsilon$ is the modified Bessel function of the second kind. The smoothness of this kernel function is determined by the value of $\upsilon$. If $\lfloor\upsilon + d/2\rfloor >d/2$ (which holds if $\upsilon \geq 1$), the reproducing kernel Hilbert space generated by this kernel function is equal to the (fractional) Sobolev space $H^{\upsilon +d/2}$ \citep[see Corollary 1 of ][]{tuo2014efficient}. Using this equivalence and the covering number of the Sobolev space \citep{edmunds2008function}, we obtain, for $\upsilon\geq 1$, the $L_\infty$ covering numbers of the balls of the reproducing kernel Hilbert space generated by Mat\'{e}rn kernel $\Phi_{\upsilon,\phi}$ are given by
\begin{eqnarray*}
\log N(\epsilon,\mathcal{N}_{\Phi_{\upsilon,\phi}}(\mathcal{X},\rho),L_\infty(\mathcal{X}))\leq \left(\frac{C_\mathcal{X} \rho}{\epsilon}\right)^{\frac{d}{\upsilon+d/2}},
\end{eqnarray*}
for a constant $C_\mathcal{X}$ depends on $\mathcal{X}$ only.
Condition~3 is clearly satisfied by this kernel with
$\nu = \upsilon + d/2$.
Native spaces generated by some other kernels may also admit covering number bound as (\ref{entropy}), e.g., the thin-plate spline and polyharmonic spline kernels \citep{duchon1977splines}. The native spaces generated by such kernels are equivalent to certain Sobolev spaces; see \cite{wendland2005scattered} for details.

We also remark that Condition~3 is satisfied for the Gaussian kernels as well. In fact, there is a much tighter entropy bound for the Gaussian case \citep{zhou2002covering}, which can yield an even faster rate of convergence. But to save space, we do not pursue this in the paper.

In this work, we do not require a specific type of emulators. Users can choose their favorite emulators (such as Gaussian process regression or polynomial models) provided that they can well approximate the underlying computer response function. Condition 4 is an assumption regarding the approximation error.

{\bf Condition 4.}
The sequence of emulators is chosen to satisfy
\begin{eqnarray}
\| \hat{\bm y}^s_n-\bm y^s\|_{C^1(\mathcal{X}\times\Theta)}=O(n^{-\frac{\nu}{2\nu+ d}}),\label{EmuCond1}
\end{eqnarray}
where $\|\bm f\|_{C^1(\Gamma)}:=\max_{j k} \sup_{x\in\Gamma}|\frac{\partial f_k}{\partial x_j}(x)|$.

Condition 4 assumes that the magnitude of the emulation error is no bigger than that of the error of the nonparametric regression.
In the case of cheap codes, $\hat{\bm y}^s_n = \bm y^s$, and thus Condition~4 automatically satisfied. In the case of expensive codes,
Condition~4 is a requirement on the rate of approximation for the emulator. The approximation error of the emulator depends on how the emulator is constructed. For commonly used emulators, their rates of convergence are available in the numerical analysis literature. For instance, the error bound for the radial basis function interpolation can be found in \cite{wendland2005scattered}. Note that the input dimension of the emulator is $d+q$ because it has both control variables and calibration parameters. 
Suppose the smoothness of emulator is $\nu'$, then a typical rate of convergence for the emulator constructed by $m$ computer outputs is $O(m^{-\nu'/(d+q)})$. Therefore, to ensure Condition 5, $m$ should be at least with the order of magnitude $O(n^{\frac{\nu}{\nu'}\frac{d+q}{2\nu+d}})$. We believe this condition can be easily achieved by choosing the sample size for the computer experiment to be much larger than $n$, which is feasible because each run of the computer codes should be much less costly than the corresponding physical experiment.

Corresponding to the design points of the physical experiment, define an empirical (semi-)norm as $\|f\|_n=\{\sum_{i=1}^n f^2(\bm x_i)/n\}^{1/2}$ for any function $f$ on $\mathcal{X}$. To incorporate the multivariate response, we extend the notion of native norm to a vector-valued function. For $\bm f=(f_1,\ldots,f_q)$, define $\|\bm f\|_\mathcal{N}$ as the Euclidean norm of $(\|f_1\|_\mathcal{N},\ldots,\|f_q\|_{\mathcal{N}})$. Similarly, we can also define the $L_2$ norm as well as the empirical norm for a vector-valued function.
	
\begin{theorem}\label{Th:prediction}
		Suppose that Conditions 1-4 are fulfilled. In addition, we assume
		\begin{eqnarray}\label{preservation}
		\|\hat{\bm y}^s_n(\bm x,\bm \theta(\bm x))\|_{\mathcal{N}}\leq C_1 \|\bm \theta\|_{\mathcal{N}}+C_2
		\end{eqnarray}
		for all $\bm \theta(\cdot)\in\mathcal{N}^{\Theta}$ and some constants $C_1,C_2>0$. Then if $\lambda_n^{-1}=o_p(n^{1/2})$ and $\lambda_n=o_p(1)$, we have $\mathrm{PE}(\hat{\bm \theta}_n)- \mathrm{PE}(\bm \theta^*) =O_p(\lambda_n)$.
\end{theorem}
	
Condition (\ref{preservation}) requires that $\hat{\bm y}^s_n(\bm x,\bm \theta(\bm x))$ preserves the smoothness of $\mathcal{N}$, which is generally true if $\hat{\bm y}^s_n$ is smooth enough. For example, if $\bm x$ and $\bm\theta$ are one-dimensional, $\Theta$ is compact, and $\mathcal{N}$ is equivalent to the Sobolev space $H^{k}$ with $k\in \mathbb{N}^+$, then according to Proposition 1.4.8 of \cite{danchin2005fourier}, (\ref{preservation}) holds if $\hat{y}^s_n$ is $k+1$ times continuously differentiable.
	
In Theorem \ref{Th:prediction}, we consider the random design case where the design points follow the uniform distribution. The fixed design is also commonly used in practical situations, where the design points are chosen in a deterministic way. Our next result extends Theorem~1 to the fixed design case. 

{\bf Condition $1'$.}
The design points $\{\bm x_1,\ldots,\bm x_n\}$ are deterministic and satisfying 
\begin{eqnarray}
\|f\|_{L_2(\mathcal{X})}\lesssim \|f\|_n+n^{-\frac{\nu}{2\nu+d}}\|f\|_{\mathcal{N}},
\label{integration}
\end{eqnarray}
for all $f\in\mathcal{N}$.
The noises $\bm e_i$'s are independent from a sub-Gaussian distribution with mean zero and sub-Gaussian parameter $\sigma$, for $i=1,\ldots,n$.

The condition (\ref{integration}) is generally a mild condition.
It holds for a broad class of design schemes called \textit{quasi-uniform designs}, which covers many commonly used space-filling designs. For a design $\mathcal{D}_n=\{\bm x_1,\ldots\bm x_n\}$, define its fill distance as
\begin{eqnarray*}
	h(\mathcal{D}_n):=\max_{\bm x\in\mathcal{X}}\min_{\bm x_i\in\mathcal{D}_n}\|\bm x-\bm x_i\|,
\end{eqnarray*}
and separation distance as
\begin{eqnarray*}
	q(\mathcal{D}_n):=\min_{\bm x_i,\bm x_j\in\mathcal{D}_n}\|\bm x_i-\bm x_j\|.
\end{eqnarray*}
Note that $h(\mathcal{D}_n)$ and $q(\mathcal{D}_n)$ are also commonly used criteria to measure the space-filling property of a design: the design minimizing $h(\mathcal{D}_n)$ is known as the \textit{minimax design} and the design maximizing $q(\mathcal{D}_n)$ is known as the \textit{maximin design}. See \cite{johnson1990minimax}. A sequence of designs $\{\mathcal{D}_n\}$ is called quasi-uniform if $h(\mathcal{D}_n)/q(\mathcal{D}_n)$ is bounded. \cite{utreras1988convergence} proved that (\ref{integration}) is ensured if the design scheme is quasi-uniform.

\begin{theorem}\label{thdeterministic1}
	Assume that Condition $1'$, Conditions 2--4 and (\ref{preservation}) are satisfied.
	 Then if $\lambda_n^{-1}=o_p(n^{1/2})$ and $\lambda_n=o_p(1)$, we have 
$\mathrm{PE}(\hat{\bm \theta}_n)- \mathrm{PE}(\bm \theta^*) =O_p(\lambda_n)$. 
\end{theorem}

\subsection{Consistency of calibration function}\label{sec_estimation}

Now we study the consistency of the empirical calibration function $\hat{\bm \theta}_n$ in estimating the optimal calibration function $\bm \theta^* $, provided the latter is uniquely defined. The required identifiability condition for $\bm \theta^*$ is specified below. 

{\bf Condition 5.}
The optimal calibration function $\bm\theta^*(x)$ exists. There exists $\omega_0>0$, such that
\begin{eqnarray}
|\bm \zeta(\bm x)-\bm y^s(\bm x,\bm \theta)|^2-|\bm \zeta(\bm x)-\bm y^s(\bm x,\bm \theta^*(\bm x))|^2\geq \omega_0 |\bm \theta-\bm \theta^*(\bm x)|^2,\label{omega0}
\end{eqnarray}
for all $\bm \theta\in\Theta$ and $\bm x\in\mathcal{X}$.

Inequality~\eqref{omega0} suggests that $|\bm \zeta(\bm x)-\bm y^s(\bm x,\cdot)|^2$ is bounded below locally by a quadratic function around the optimal calibration parameter $\bm \theta^*(\bm x)$. It also implies that $\bm \theta^*(\cdot)$ is uniquely defined. 
A sufficient condition for \eqref{omega0} is that $\Theta$ is a compact set, $\bm y^s\in C^2(\mathcal{X}\times\Theta)$ and
\begin{eqnarray}
\inf_{\bm x\in\mathcal{X}}\lambda_{min}\left(\frac{\partial^2 l_{\bm x}(\bm \theta)}{\partial \bm  \theta \partial \bm \theta^T} \bigg\vert_ {{\bm \theta} = \bm \theta^*(\bm x)}  \right)=\lambda_0>0,\label{Hessian}
\end{eqnarray}
where $l_{\bm x}(\theta)=|\bm \zeta(\bm x)-\bm y^s(\bm x, \bm \theta)|^2$ and $\lambda_{min}(A)$ denotes the smallest eigenvalue of the matrix $A$. To see this, we apply the Taylor expansion to obtain that for any $\bm \theta\in\Theta$,
\begin{eqnarray*}
	l_{\bm x}(\bm \theta)-l_{\bm x}( \bm \theta^*(\bm x))=(\bm \theta-\bm \theta^*(\bm x))^T\frac{\partial^2 l_{\bm x}(\bm \theta)}{\partial \bm  \theta \partial \bm \theta^T} \bigg\vert_ {{\bm \theta} = \bm \theta'}  (\bm \theta-\bm \theta^*(\bm x)),
\end{eqnarray*}
where $\bm \theta'$ lies between $\bm \theta$ and $\bm \theta^*(\bm x)$. 
Let $\bm A \ge \bm B$ mean $\bm A - \bm B$ is positive semi-definite for square matrices $\bm A, \bm B$. Because of the condition $\bm y^s\in C^2(\mathcal{X}\times\Theta)$ and (\ref{Hessian}), we can find $\delta>0$ so that
\begin{eqnarray}
\frac{\partial^2 l_{\bm x}(\bm \theta) }{\partial \bm  \theta \partial \bm \theta^T} \bigg\vert_ {{\bm \theta} = \bm \theta'} 
\geq \frac{1}{2}\frac{\partial^2 l_{\bm x}(\bm \theta) }{\partial \bm  \theta \partial \bm \theta^T} \bigg\vert_ {{\bm \theta} = \bm \theta^*(\bm x)}   \geq \frac{\lambda_0}{2} {\bm I}, \label{eigen}
\end{eqnarray}
for all $|\bm \theta-\bm \theta^*(\bm x)|\leq \delta$ and $\bm x\in\mathcal{X}$. This implies
\begin{eqnarray}
l_{\bm x}(\bm \theta)-l_{\bm x}(\bm \theta^*(\bm x))\geq \lambda_0 |\bm \theta-\bm \theta^*(\bm x)|^2/2, \label{part1}
\end{eqnarray}
for all $|\bm \theta-\bm \theta^*(\bm x)|\leq \delta$.
On the other hand, because $\theta^*$ is the unique optimal (minimum distance) calibration function, for $|\bm \theta-\bm \theta^*(\bm x)|> \delta$, there exists a constant $\gamma>0$ so that
\begin{eqnarray}
l_{\bm x}(\bm \theta)-l_{\bm x}(\bm \theta^*(\bm x))\geq \gamma\geq \gamma |\bm \theta- \bm \theta^*|^2/D(\Theta)^2,\label{part2}
\end{eqnarray}
where $D(\Theta)$ denotes the diameter of $\Theta$. Thus (\ref{omega0}) is ensured by combining (\ref{part1}) and (\ref{part2}). Note that (\ref{Hessian}) is a mild condition analogous to a standard one in parametric statistical inference: the Fisher information matrix is non-singular (or positive definite).

The main result for the calibration consistency is Theorem \ref{ThCR}.  The rate of convergence here can be much faster than that in Theorem \ref{Th:prediction}, and attains the minimax rate for nonparametric regression \citep{stone1982optimal}.
In Theorem \ref{ThCR}, the smoothing parameter $\lambda_n$ can be random but subject to certain order of magnitude conditions.

\begin{theorem}\label{ThCR}
Assume Conditions 1--5 are satisfied. Assume further that the sequence $\{\lambda_n\}$ is chosen to satisfy  $\lambda_n\sim n^{-\frac{2\nu}{2\nu+ d}}$, that is,
$\lambda_n=O_p(n^{-\frac{2\nu}{2\nu+ d}})$ and $\lambda^{-1}_n=O_p(n^{\frac{2\nu}{2\nu+ d}})$.
Then $\|\hat{\bm \theta}_n-\bm \theta^*\|_{L_2(\mathcal{X})}=O_p(n^{-\frac{\nu}{2\nu+ d}})$ and $\|\hat{\bm \theta}_n\|_{\mathcal{N}}=O_p(1)$.
\end{theorem}

Similar to Theorem \ref{thdeterministic1}, we have a fixed-design version of the calibration convergence theorem, given by Theorem \ref{thdeterministic}.

\begin{theorem}\label{thdeterministic}
Assume that Condition $1'$ and Conditions 2--5 are satisfied.
Assume further that the sequence $\{\lambda_n\}$ is chosen to satisfy $\lambda_n\sim n^{-\frac{2\nu}{2\nu+d}}$,
$\lambda_n=O(n^{-\frac{2\nu}{2\nu+ d}})$ and $\lambda^{-1}_n=O(n^{\frac{2\nu}{2\nu+ d}})$.
Then $\|\hat{\bm \theta}_n-\bm \theta^*\|_{L_2(\mathcal{X})}=O_p(n^{-\frac{\nu}{2\nu+ d}})$ and $\|\hat{\bm \theta}_n\|_{\mathcal{N}}=O_p(1)$.
\end{theorem}

{\bf Remark.} Conditions~1 and $1'$ assume sub-Gaussion noises but this condition can be relaxed.  By applying an adaptive truncation argument using Bernstein's inequality \citep{van2000empirical}, it can be proved that Theorems \ref{Th:prediction}-\ref{thdeterministic} hold for sub-exponential $e_i$'s, i.e., $E[e^{c |e_i|}]<+\infty$ for some $c>0$.

{\bf Remark.} The method in this paper is inspired by smoothing splines, a widely used method for nonparametric regression. However, this paper addresses the problem of computer model calibration, which is a problem different from nonparametric regression. For readers interested in comparing results in this section with asymptotic results of smoothing splines, we refer to Gu (2013) and van der Geer (2000) for results of rates of convergence for smoothing splines.


\section{Computation}~\label{sec:computation}
This section develops an algorithm to solve the minimization problem~\eqref{eq:LC_gen} and a method for penalty parameter selection. The same algorithm is applicable to solve \eqref{eq:LC_gen_exp}. For simplicity, we  drop the constraint $\bm \theta(\cdot) \in \Theta$ since we can monitor the steps of the iterative algorithm to ensure that the iteration steps will not lead to a point outside the feasible region.  Moreover, we focus on the case of one-dimensional response (i.e., $r=1$). Extension of the algorithm to general $r$ is straightforward with only some notational complications.  After these simplifications, the problem ~\eqref{eq:LC_gen} becomes
\begin{equation} \label{eqn:newobj0}
\min_{\bm\theta(\cdot) = (\theta_1(\cdot), \dots, \theta_q(\cdot))^T }  
\sum_{i=1}^n \{y_i^p - y^s(\bm x_i, \bm \theta(\bm x_i))\}^2 + n\lambda 
\sum_{j=1}^q\Vert\theta_j \Vert^2_{\mathcal{N}_\Phi}.
\end{equation}
Here $y_i^p$ and $y^s(\bm x_i, \bm \theta(\bm x_i))$ are scalers. 

We first reduce the optimization over the infinite-dimensional native space $\mathcal{N}_{\Phi}$ to an optimization over a finite-dimensional space. Let $\mathcal{N}_0$ be the finite-dimensional null space of $\mathcal{N}_{\Phi}(\mathcal{X})$, i.e. $\mathcal{N}_0=\{f:\|f\|_{\mathcal{N}_{\Phi}}=0\}$. Let $\{v_{1},v_{2}\dots,v_{k}\}$ be a set of basis functions for $\mathcal{N}_0$. The generalized representer theorem~\citep[e.g.,][]{scholkopf2001generalized} implies that a solution to \eqref{eq:LC_gen} has the basis expansion
\begin{equation}\label{eqn:newobjSolution}
\theta_j(\bm x) = v_j(\bm x) + \phi_j(\bm x) := \sum_{s=1}^k \alpha_{js} v_{s}(\bm x) +
\sum_{i=1}^n\beta_{ji} \Phi(\bm x, \bm  x_i)\, .
\end{equation}
for $j=1,2,\cdots, q$.
The solution is a sum of $v_j(\bm x)$ in the null space and $\phi_j(\bm x)$ in the reproducing kernel Hilbert space generated by the kernel function $\Phi$. 
With the representation in \eqref{eqn:newobjSolution}, the quasi-Newton algorithm~\citep{byrd1989tool} can be used to solve \eqref{eqn:newobj0} and worked well in our numerical studies. 

The penalty parameter $\lambda$ can be selected in a similar manner as in the smoothing spline literature, by minimizing the generalized cross-validation or GCV criterion \citep{GolubHeathWahba1979, CravenWahba1979}. To develop an appropriate GCV criterion in our context, we extend the derivations in Chapter 3 of \citet{gu2013smoothing}. 

We expand the computer model $y^s(\bm x,\bm\theta(\bm x))$  by its first-order Taylor expansion at the  
 optimal value  $\hat{\bm \theta}(\bm x)$,  
\begin{equation}\label{eq:approx-model}
y^s(\bm x_i,\bm \theta(\bm x_i) )\approx 
y^s(x_i, \hat{\bm \theta}(\bm x_i) ) +
\sum_{j=1}^{q}
\frac{\partial}{\partial \theta_j} y^s(\bm x_i, \hat{\bm \theta}(\bm x_i) )
\times \left(\theta_j(\bm x_i) -\hat{\theta}_j(\bm x_i)\right).
\end{equation}
Plug this into the original objective function, we get
\begin{equation*}
\min_{\theta(\cdot)}\, \sum_{i=1}^n \biggl\{y_i^p - 
y^s(\bm x_i, \hat{\bm\theta}(\bm x_i) ) -
\sum_{j=1}^q
\frac{\partial}{\partial \theta_j} y^s(\bm x_i, \hat{\bm \theta}(\bm x_i) )
\times\left(\theta_j(\bm x_i) -\hat{\theta}_j(\bm x_i)\right)
\biggr\}^2 + n\lambda \sum_{j=1}^q\Vert
\theta_j \Vert^2_{\mathcal{N}_\Phi}
\end{equation*}
Define $w_{ij} =  \frac{\partial}{\partial \theta_j} y^s(\bm x_i,
\hat{\bm \theta}(\bm x_i) ) $ and
$ \bar{y}_i =  y_i^p -  y^s(x_i, \hat{\bm \theta}(x_i) ) 
+ \sum_{j=1}^{q} w_{ij} \hat{\theta}_j(\bm x_i) $.
The original problem is then reduced to a penalized regression problem, 
\begin{equation} \label{eqn:newobj1}
\min_{\theta(\cdot)}\, \sum_{i=1}^n \biggl\{ \bar{y}_i - 
\sum_{j=1}^{q} w_{ij}\theta_j(x_i)  
\biggr\}^2 + n\lambda  \sum_{j=1}^{q}\Vert \theta_j \Vert^2_{\mathcal{N}_\Phi}.
\end{equation}
 Let 
$\valpha_j = (\alpha_{j1}, \dots, \alpha_{jk})^T$ and 
$\vbeta_j = (\beta_{j1}, \dots, \beta_{jn})^T$. The reproducing property of the kernel function $\Phi$ implies that 
$\Vert \theta_j \Vert^2_{\mathcal{N}_\Phi} =  \vbeta^T_j \bPhi\vbeta_j$, 
where $\bPhi = (\Phi(x_i, x_j))_{n\times n}$.
Let $\bar{\bm y} = (\bar{y}_1, \cdots, \bar{y}_n)^T$, $\bV = (v_s(\bm x_i))_{n\times q}$, and 
the diagonal weight matrix $\bW_j = \mathrm{diag}(w_{1j}, \cdots, w_{nj})$.
The optimization problem~(\ref{eqn:newobj1}) is expressed in the matrix notation as
\begin{equation} \label{eqn:newobj2}
\min_{\valpha_j, \vbeta_j}\,  \biggl\Vert 
\bar{\bm y} - \sum_{j=1}^{q} 
\bW_j\left(\bV\valpha_j + \bPhi\vbeta_j \right) \biggr\Vert_2^2 
+ n\lambda \sum_{j=1}^q\vbeta^T_j\bPhi\vbeta_j \, ,
\end{equation}
which in turn is equivalent to
\begin{equation} \label{eqn:newobj3}
\min_{\valpha_j, \vbeta_{jw}}\,  \biggl\Vert 
\bar{\bm y} - \sum_{j=1}^{q} 
\left(\bV_{jw}\valpha_{j} + \bPhi_{jw}\vbeta_{jw} \right) \biggr\Vert_2^2 
+ n\lambda \sum_{j=1}^q\vbeta^T_{jw}\bPhi_{jw}\vbeta_{jw} \, .
\end{equation}
where $\bV_{jw} = \bW_j \bV$, $\bPhi_{jw} = \bW_j\bPhi\bW_j$, and $\vbeta_{jw} = \bW^{-1}_j\vbeta_j$.  

For the objective function in (\ref{eqn:newobj3}), take first order derivative
with respect to $\valpha_j$ and $\vbeta_{jw}$ and set them to zero.  We
find that the optimal solution must satisfy 
\begin{eqnarray}
\bV_{jw}^T  \biggl\{\bar{\bm y} - 
\sum_{j=1}^{q} 
\left(\bV_{jw}\valpha_{j} + \bPhi_{jw}\vbeta_{jw} \right)\biggr\}
&=&  \bm 0 \label{eqn:solveAB1}\, ,  \\
\bPhi_{jw}  \biggl\{\bar{\bm y} - 
\sum_{j=1}^{q} 
\left(\bV_{jw}\valpha_{j} + \bPhi_{jw}\vbeta_{jw} \right)\biggr\} 
- n\lambda \bPhi_{jw}\vbeta_{jw}
  &= & \bm 0 \label{eqn:solveAB2} \, .
\end{eqnarray}
If $\bPhi_w$ is singular, the above equation may have multiple
solutions.  It is easy to see equation (\ref{eqn:solveAB1}) and
(\ref{eqn:solveAB2}) are satisfied by the solution of
\begin{eqnarray}
\bV_{kw}^T  \vbeta_{jw} &=& \bm 0\, , \label{eqn:solveAB3} \\
\bar{\bm y} - \sum_{j=1}^{q} 
\left(\bV_{jw}\valpha_{j} + \bPhi_{jw}\vbeta_{jw} \right)
- n\lambda \vbeta_{jw}
&= & \bm 0\, , \label{eqn:solveAB4}
\end{eqnarray}
for any $j,k = 1,2,\cdots, q$ and the solution is unique. In fact, pre-multiply
 (\ref{eqn:solveAB4}) by $\bPhi_{jw}$, we get equation (\ref{eqn:solveAB2}).
In addition, equation (\ref{eqn:solveAB3}) for $k = j$ and equation (\ref{eqn:solveAB4})
together imply (\ref{eqn:solveAB1}).

These equations can be arranged into a more compact form. 
Let $\valpha = (\valpha_1^T, \valpha_2^T, \cdots, \valpha_q^T)^T$, 
and $\vbeta_w = (\vbeta_{1w}^T, \vbeta_{2w}^T, \cdots, \vbeta_{qw}^T)^T$. We 
also let $\bar{\bm y}$ be repeated $q$ times, and the resulting vector is denoted as
$\bm Y = (\bar{\bm y}^T, \cdots, \bar{\bm y}^T)^T\in \mathbf{R}^{qn}$.
Similarly, we define
\begin{equation*}
	\bV_w = \begin{pmatrix}
	\bV_{1w} & \bV_{2w} & \cdots & \bV_{qw} \\
	\bV_{1w} & \bV_{2w} & \cdots & \bV_{qw} \\
	\vdots & \vdots &  & \vdots \\
	\bV_{1w} & \bV_{2w} & \cdots & \bV_{qw} \\
	\end{pmatrix},
	\quad
		\bPhi_w = \begin{pmatrix}
	\bPhi_{1w} & \bPhi_{2w} & \cdots & \bPhi_{qw} \\
	\bPhi_{1w} & \bPhi_{2w}  & \cdots & \bPhi_{qw} \\
	\vdots & \vdots &  & \vdots \\
	\bPhi_{1w} & \bPhi_{2w} & \cdots & \bPhi_{qw} \\
	\end{pmatrix}.
\end{equation*}
where the row blocks simply repeat the first row block $q$ times. 
It follows that equations (\ref{eqn:solveAB3})
and (\ref{eqn:solveAB4}) can be presented as
\begin{eqnarray}
\bV_w^T  \vbeta_{w} &=& \bm 0\, , \label{eqn:solveAB5} \\
\bY -  \bV_w\valpha - (\bPhi_w + n\lambda \bI)\vbeta_{w}
&= & \bm 0\, , \label{eqn:solveAB6}
\end{eqnarray}
This requires $\vbeta_w$ to be orthogonal to the column space of
 $\bV_w$. 
Suppose the full QR decomposition of 
$\bV_w $ is $\bV_w = (\bF_1\, \bF_2)
(\mathbf{R}^T \mathbf{0}^T)^T$, then we have $\bF_1^T\vbeta_w = 0$ and
$\vbeta_w = \bF_2\bF_2^T \vbeta_w$. Pre-multiply (\ref{eqn:solveAB6}) by
$\bF_2^T$ and $\bF_1^T$ separately, the solution to (\ref{eqn:solveAB5}) and
(\ref{eqn:solveAB6}) can be found as
\begin{eqnarray}
\hat{\vbeta}_w &=& \bF_2\left(\bF_2^T\bPhi_w\bF_2 + n\lambda
                  \bI\right)^{-1}\bF_2^T \bY \, ,\label{eqn:get-beta}\\
\hat{\valpha} &= & \mathbf{R}^{-1}\left( \bF_1^T\bY -
            \bF_1^T\bPhi_w\hat{\vbeta}_w\right)\, .\label{eqn:get-alpha} 
\end{eqnarray}
Plug the above solution into \eqref{eqn:solveAB6} and with some calculation, 
we obtain that the estimated $\bY$ is 
$\hat{\bY} = \bV_w\hat{\valpha} + \bPhi_w\hat{\vbeta}_w =
\bA(\lambda) \bY$, where
$ \bA(\lambda) = \bI - n\lambda \bF_2(\bF_2^T\bPhi_w\bF_2 + 
n\lambda  \bI)^{-1} \bF_2^T$.

The smoothing matrix $\bA(\lambda)$ transforms $\bY$ to $\hat{\bY}= \bA(\lambda) \bY$. This connection
suggests the following generalized cross-validation (GCV) criterion
\begin{equation}\label{eqn:gcvscore}
{\rm GCV}(\lambda) = \frac{(qn)^{-1} \bY^T (\bI - \bA(\lambda))^2\bY}{
 	 \{(qn)^{-1} \mathrm{tr}(\bI - \bA(\lambda) )\}^2 } \, .
\end{equation}
The term $q$ appears in the above equation because the whole set of observations is repeated $q$ times in our derivation.
This GCV is employed to select the penalty parameter $\lambda$.

\section{Uncertainty Quantification}~\label{sec:UQ}
The uncertainty of estimation of the calibration function and prediction comes from two sources: the physical experiments and the computer experiments. We develop an uncertainty quantification method that considers only the former source of uncertainty, which is usually the dominant source of uncertainty. This method is adequate for use in the case of cheap codes, and may under-estimate the uncertainty in the case of expensive codes. As in the previous section, our presentation focuses on the case of one-dimensional responses (i.e., $r=1$).

Adopting a strategy typically used in the smoothing spline regression literature \citep[e.g.,][]{wahba1990spline,gu2013smoothing}, we develop a confidence interval for $\hat{\bm \theta}(\bm x)$ at an arbitrary $\bm x$ via considering the penalized regression~(\ref{eqn:newobj1}) in the Bayesian framework. Let $\sigma^2_e$ denote the variance of the measurement noise $e_i$ in data model \eqref{eq:data-model}. The
objective function of~(\ref{eqn:newobj1}) is proportional to
$$
\sum_{i=1}^n -\frac{1}{2\sigma^2_e} \biggl\{ \bar{y}_i - 
\sum_{j=1}^{q} w_{ij}\theta_j(x_i)  
\biggr\}^2 -\frac{n\lambda}{2\sigma^2_e}
\sum_{j=1}^{q}\Vert \theta_j \Vert^2_{\mathcal{N}_\Phi}
$$
Taking exponential of the above, we arrive at
\begin{equation}\label{eqn:newjoint} 
	\prod_{i=1}^n \exp\biggl(
	-\frac{1}{2\sigma^2_e} \biggl\{ \bar{y}_i - 
	\sum_{j=1}^{q} w_{ij}\theta_j(x_i)  
	\biggr\}^2\biggr) \times 
	\prod_{j=1}^q \exp\biggl( -\frac{n\lambda}{2\sigma^2_e}
	\Vert \theta_j \Vert^2_{\mathcal{N}_\Phi} \biggr)\, .
\end{equation}

Firstly,  recall that $w_{ij} =  \frac{\partial}{\partial \theta_j} y^s(\bm x_i,
\hat{\bm \theta}(\bm x_i) )$ and
$\bar{y}_i =  y_i^p -  y^s(x_i, \hat{\bm \theta}(x_i) ) 
+ \sum_{j=1}^{q} w_{ij} \hat{\theta}_j(\bm x_i) $.
When the estimated $\hat{\bm \theta}$ is close to the population optimal 
$\bm \theta^*$,  the distribution of $\bar{y}_i$ is close to that of
\begin{equation}
	y_i^p -  y^s(\bm x_i, \bm \theta^*(\bm x_i) ) 
	+ \sum_{j=1}^{q} w_{ij} \theta^*_j(\bm x_i)
	 = \sum_{j=1}^{q} w_{ij} \theta^*_j(\bm x_i) + \epsilon_i.
\end{equation}
In other words, the term $ \bar{y}_i $ approximately follows a normal distribution with 
mean $\sum_{j=1}^{q} w_{ij}\theta_j^*(x_i)$ and variance $\sigma^2$, which is the first term
in \eqref{eqn:newjoint}.

Secondly, the penalty on $\theta_j(\cdot)$ is equivalent to
imposing a prior on the native space $\mathcal{N}_\Phi$ corresponding to  kernel $\Phi$.  
Consider the decomposition (\ref{eqn:newobjSolution}),
for the function $\phi_j(\cdot)$ in the reproducing kernel Hilbert space, one can think it follows a Gaussian process prior
$\phi_j(\cdot)\sim\mathcal{GP}\big(0, \frac{\sigma^2_e}{n\lambda}\Phi(\cdot, \cdot)\big)$. 
Lastly, we can impose an additional prior on the coefficients
$\alpha_{js} \sim N(0, \rho\, \sigma^2_e/(n\lambda))$. The contant $\rho$ takes a large value, resulting in an
approximately non-informative prior on the null space. 
Combining the above leads to the joint distribution (\ref{eqn:newjoint})
when $\rho$ approaches infinity. 

At an arbitrary location $\bm x$, define 
$\bv = (v_1(\bm x), \cdots, v_s(\bm  x))^T$ and $\bphi = (\bPhi(\bm x,\bm  x_1) ,\cdots,
\bPhi(\bm x, \bm  x_n))^T$. The joint distribution of $\theta_j(\bm  x)$ and
$\bar{\bm y}$ is a normal distribution 
$$\mathcal{N}\left(
\bm 0, \frac{\sigma^2_e}{n\lambda}\begin{pmatrix}
	\sigma_{11} & \mathbf{\Sigma}_{21}^T \\
	\mathbf{\Sigma}_{21} & \mathbf{\Sigma}_{22}
\end{pmatrix}\right),$$
where
$\sigma_{11} = \Phi(\bm x,\bm  x) + \rho \bv^T \bv$, 
$\mathbf{\Sigma}_{21} = \bW_j\bphi + \rho \bV_{jw}\bv$, and 
$\mathbf{\Sigma}_{22} = \sum_{j=1}^q (\bPhi_{jw}+ \rho\bV_{jw}\bV_{jw}^T) + n\lambda\bI$. 
It follows that the conditional variance of $\theta_j(\bm  x)$
given $\bar{\bm y}$ is 
\begin{equation} \label{eq:cond-var}
	\hat{\sigma}_{\theta_j}^2(\bm x) = 
	\frac{\sigma^2_e}{n\lambda}\left(
	\sigma_{11}  - \mathbf{\Sigma}_{21}^T
	\mathbf{\Sigma}_{22}^{-1}\mathbf{\Sigma}_{21}
	\right)\,.
\end{equation}

To make use of this distribution result, we need an estimate of $\sigma^2_e$.
Based on the converted regression problem (\ref{eqn:newobj3}), we adopt
the following estimator \citep[equation 3.26 in][]{gu2013smoothing}
\begin{equation}\label{eqn:sigmaEst}
	\hat{\sigma}^2_e = \frac{\bY_w^T (\bI - \bA(\lambda))^2
		\bY_w}{ \mathrm{tr}(\bI - \bA(\lambda) ) }\, ,
\end{equation}
and $\lambda$ is chosen by minimizing the GCV criterion (\ref{eqn:gcvscore}). 

Putting the above together, the $(1-\alpha)\times 100\%$ confidence interval for $\theta_j(\bm x)$ is 
\begin{equation}\label{eq:conf-parameter}
	(\hat{\theta}_j(\bm x) - z_{\alpha/2}\hat{\sigma}_{\theta_j}(\bm x), \;
	\hat{\theta}_j(\bm x) + z_{\alpha/2}\hat{\sigma}_{\theta_j}(\bm x)),
\end{equation}
where $z_{\alpha}$ is the upper $\alpha$ quantile of the standard normal
distribution and $\hat{\sigma}_{\theta_j}(\bm x)$ is calculated using the expression \eqref{eq:cond-var} with $\sigma^2_e$ estimated by \eqref{eqn:sigmaEst} and $\lambda$ selected by the GCV.
The confidence intervals for functional calibration parameters given in \eqref{eq:conf-parameter} only make sense when the optimal calibration functions are identifiable.

While the conditional variance of $\bm \theta(\bm x)$ given $\bar{\bm y}$ can be computed as~\eqref{eq:conf-parameter}, 
the conditional variance of $y^s(\bm x, \bm \theta(\bm x))$ given $\bar{\bm y}$, denoted as $\sigma_{y^s}^2(\bm x)$,  
is easily obtained by the delta method. In particular, when the gradient is 
	non-zero, i.e.  $\frac{\partial}{\partial\bm \theta} y^s(\bm x,
	\hat{\bm \theta}(\bm x) )\neq \mathbf{0} $, we can compute
\begin{equation}\label{eq:var-prediction}
	\sigma^2_{y^s}(\bm x) =
	\frac{\sigma^2_e}{n\lambda}\left(\Lambda_{11} -  \mathbf{\Lambda}_{21}^T 
	\mathbf{\Lambda}_{22}^{-1}
	\mathbf{\Lambda}_{21}\right),
\end{equation}
where 
$\Lambda_{11} =\sum_{j=1}^q w_j^2 (\Phi(\bm x,\bm  x) + \rho \bv^T \bv)$,
$\mathbf{\Lambda}_{21} = \sum_{j=1}^q w_j (\bW_j\bphi + \rho \bV_{jw}\bv)$,
$\mathbf{\Lambda}_{22} = \sum_{j=1}^q (\bPhi_{jw}+ \rho\bV_{jw}\bV_{jw}^T) + n\lambda \bI$, 
and $w_j = \frac{\partial}{\partial \theta_j} y^s(\bm x,
\hat{\bm \theta}(\bm x) )$.
This expression can be equivalently obtained from the the covariance matrix of 
$y^s(\bm x, \bm \theta(\bm x))$ and $\bar{\bm y}$, which is computed with the aid of \eqref{eq:approx-model} to be
$$
\frac{\sigma^2_e}{n\lambda} \mathbf{\Lambda} = 
\frac{\sigma^2_e}{n\lambda} \begin{pmatrix}
	\Lambda_{11} & \mathbf{\Lambda}_{21}^T \\
	\mathbf{\Lambda}_{21} & \mathbf{\Lambda}_{22} 
\end{pmatrix}.
$$

The $(1-\alpha)\times 100\%$  confidence interval for the computer model response $y^s(\bm x, \bm \theta^*(\bm x))$ at location $\bm x$ is
\begin{equation}\label{eq:conf-int-resp}
(y^s(\bm x,\hat{\bm\theta} (\bm x)) - z_{\alpha/2}\hat{\sigma}_{y^s}(\bm x)), \;
y^s(\bm x,\hat{\bm\theta}(\bm x)) + z_{\alpha/2}\hat{\sigma}_{y^s}(\bm x))),
\end{equation}
where $\hat{\sigma}_{y^s}(\bm x)$ is calculated using \eqref{eq:var-prediction} with $\sigma^2_e$ estimated by using \eqref{eqn:sigmaEst}. This can be used as a confidence interval for the physical response $\zeta(\bm x)$. The confidence interval of the physical response given in \eqref{eq:conf-int-resp} is meaningful even when $\theta^*(\bm x)$ is unidentified, since 
the optimal prediction function $y^s(\bm x, \bm \theta^*(\bm x))$ is uniquely defined.

We expect that the confidence intervals in \eqref{eq:conf-parameter} and \eqref{eq:conf-int-resp} have the across-the-function coverage property \citep{wahba1983, nychka1988}. Rigorous asymptotic justification is left for future research. In the next section, we will empirically illustrate the performance of these intervals in a simulation study.


\section{Simulation Study}\label{sec_res}

In this section, we compare our proposed method, the nonparametric functional calibration, with the constant
calibration~\citep[abreviated as Const;][]{Tuo2014calibration}, the parametric functional calibration~\citep{Pourhabib2014bp},  the Bayesian method of \citet{brown2018nonparametric}. To set up the stage for comparison, we consider two parametric calibration models in Section~\ref{sec:parametric} and two unidentified calibration models in Section~\ref{sec:unidentified-model}. For parametric functional calibration, we consider two parametric
models. The first is the parametric exponential (Param-Exp) model, $\theta(x) = \gamma_0 \exp(\gamma_1  x)$.
The second is a quadratic model (Param-Quad)  of the form $\theta(x) =\gamma_0 + \gamma_1 x +   \gamma_2 x^2$.
For our proposed method,  the native space is chosen as the Sobolev space of order $2$. It has kernel 
$k_2(x,y) = \frac{1}{(2!)^2}B_2(x)B_2(y) - \frac{1}{4!}B_4(\vert
x-y\vert)$, where $B_m(\cdot)$ is the $m$th Bernoulli polynomial \citep[e.g., page 39 of][]{gu2013smoothing}. This kernel is commonly used in the smoothing spline literature and the solution in this space is a cubic spline function. In the following, our nonparametric functional calibration method using this kernel is denoted as RKHS-Cubic. 

In the comparative study of Section 6.2 for the cases of unidentified calibration models, we also include local approximate Gaussian process regression method \citep[abreviated as laGP;][]{gramacy2015calibrating,gramacy2016lagp}, 
which runs a Gaussian process regression on the residuals from a constant calibration model.  The laGP method is a mis-specified model in the context of Section 6.1 and thus not considered in that section where the estimation error of the calibration parameter is evaluated.

In our study, we consider both cases of cheap code and expensive code. This allows us to evaluate the impact of using the emulator on each method in the case of expensive code. In the cheap code cases, the exact computer model $y^{s}(x,\theta)$ are used for calibration. In the expensive code cases, the computer model is evaluated on a grid of size $14\times 15$ on the domain of interest and a Gaussian process emulator with the squared exponential kernel \citep[e.g., page 83,][]{rasmussen2006gaussian} is trained on the computer generated data. The output of the emulator serves as a surrogate computer model for calibration. 

\subsection{Two parametric models}~\label{sec:parametric}
For each of the two settings, one of the above two parametric calibration models (Param-Exp or Param-Quad) exactly matches the physical model for a given parameter. This allows us to compare the performance of the parametric calibration model when the model is correctly specified and also when the model is mis-specified. 
\begin{enumerate}
\item The first setting. The physical response is 
 $y^p(x) = \exp(x/10) \cos(x) + \sigma e$ for $x\in[\pi, 3\pi]$, where $\sigma=0.1$ and $e \sim N(0,1)$. The
computer model is $y^s(x, \theta) = 0.5 \exp(x/10) \cos(x) \frac{\exp(x/5)}{\theta}$ with the calibration parameter $\theta$.
The optimal calibration function is $\theta^{*}(x) = 0.5 \exp(x/5) $. The emulator is trained on the domain $[\pi,
3\pi]\times[\pi/5, 6\pi/5]$.
\item The second setting. The physical response is $y^p(x) = \cos(2x)
  \sin(x/2) +  \sigma e$ for $x\in[0.5\pi, \pi] $, where $\sigma=0.1$ and $e \sim N(0,1)$.
The computer model is 
$$y^s(x, \theta) =  \cos(2x) \sin(x/2) \exp\left( \frac{3\theta}{0.5(x-2)^2+0.5} -3\right)$$ 
with the calibration parameter $\theta$.
The optimal calibration function is $\theta^{*}(x) = 0.5 (x-2)^2 + 0.5$. The emulator is trained on the domain $[0.5\pi, \pi]\times[\pi/9, \pi/2]$.
\end{enumerate}

For both the  cheap and expensive code cases in these two settings, the
simulation is repeated 100 times. In each replication, $n=50$ sample
points are generated from on the physical response model with the design points $x_i$'s uniformly generated on the domain. 

The following metrics are employed to compare different methods.
The accuracy of the estimate $\hat{\theta}(x)$ is measured by the $L_2$-loss, 
$ \{\int_{\mathcal{X}} (\theta^{*}(x) - \hat{\theta}(x) )^2 dx\}^{1/2}$.
For each method, confidence intervals are constructed for three nominal
levels of coverage probability, $90\%$, $95\%$ and $99\%$.
Suppose the upper and lower bound of the confidence interval are
$U(x)$ and $L(x)$. Its average width across the domain $\mathcal{X}$ 
is computed as  $ \int_{\mathcal{X}}  \{U_{0.05}(x) - L_{0.05}(x)\}\, dx$.
The average coverage rate (CR) across $\mathcal{X}$ is measured as
$\int_{\mathcal{X}} 
I(U_\alpha (x) > \theta^{*}(x) > L_\alpha(x))\,dx / |\mathcal{X}|$.
The integrals are approximated by the Riemann sum with 200 equally spaced points on the domain.

\begin{table}[t]
	\centering
	\caption{Simulation 1 (Param-Exp model). Comparison of methods with the cheap
		code (CC) and expensive code (EC). The mean (and SE) of the $L_2$-loss and of the width and average coverage rate
		(CR) for the level 90\%, 95\% and 99\% confidence intervals.} \label{tbl:simulation1}
	\begin{tabular}{|c|c|c|cc|cc|cc|}
		\hline
		\multirow{2}{*}{Code} 
		&\multirow{2}{*}{Method} & \multirow{2}{*}{$L_2$-loss} & 
		\multicolumn{2}{c|}{90\%}
		& \multicolumn{2}{c|}{95\%}
		& \multicolumn{2}{c|}{99\%} \\
		& & & Width & CR & Width & CR & Width & CR\\
	\hline
CC &  Const   &  2.222 &  2.376 &  0.123 &  2.831 &  0.146 &  3.721 &  0.193 \\
&    &  (0.022) &  (0.027) &  (0.001) &  (0.032) &  (0.002) &  (0.042) &  (0.002) \\
&  Param-Exp   &  0.063 &  0.527 &  0.883 &  0.628 &  0.946 &  0.825 &  0.989 \\
&    &  (0.003) &  (0.006) &  (0.021) &  (0.007) &  (0.014) &  (0.010) &  (0.007) \\
&  Param-Quad   &  0.086 &  0.620 &  0.822 &  0.739 &  0.891 &  0.971 &  0.966 \\
&    &  (0.003) &  (0.008) &  (0.020) &  (0.009) &  (0.016) &  (0.012) &  (0.009) \\
&  RKHS-Cubic   &  0.131 &  1.074 &  0.895 &  1.280 &  0.937 &  1.682 &  0.979 \\
&    &  (0.004) &  (0.021) &  (0.012) &  (0.025) &  (0.010) &  (0.033) &  (0.005) \\
&  Baysian   &  0.157 &  1.696 &  0.960 &  2.030 &  0.979 &  2.693 &  0.994 \\
&    &  (0.005) &  (0.019) &  (0.006) &  (0.022) &  (0.004) &  (0.029) &  (0.001) \\
\hline
EC &  Const   &  2.288 &  2.550 &  0.129 &  3.038 &  0.155 &  3.993 &  0.204 \\
&    &  (0.024) &  (0.027) &  (0.002) &  (0.033) &  (0.002) &  (0.043) &  (0.002) \\
&  Param-Exp   &  0.072 &  0.535 &  0.830 &  0.637 &  0.903 &  0.837 &  0.965 \\
&    &  (0.005) &  (0.007) &  (0.024) &  (0.008) &  (0.019) &  (0.010) &  (0.012) \\
&  Param-Quad   &  0.095 &  0.643 &  0.827 &  0.766 &  0.891 &  1.007 &  0.955 \\
&    &  (0.005) &  (0.008) &  (0.021) &  (0.010) &  (0.017) &  (0.013) &  (0.011) \\
&  RKHS-Cubic   &  0.164 &  1.296 &  0.890 &  1.544 &  0.935 &  2.029 &  0.981 \\
&    &  (0.007) &  (0.030) &  (0.011) &  (0.036) &  (0.009) &  (0.047) &  (0.005) \\
&  Baysian   &  0.165 &  1.762 &  0.948 &  2.108 &  0.972 &  2.792 &  0.996 \\
&    &  (0.006) &  (0.018) &  (0.007) &  (0.022) &  (0.005) &  (0.030) &  (0.002) \\
\hline
	\end{tabular}
\end{table}

\begin{table}[t]
	\centering
	\caption{Simulation 2 (Param-Quad model). Comparison of methods with the cheap
		code (CC) and expensive code (EC). The mean (and SE) of the $L_2$-loss and of the width and average coverage rate
		(CR) for the level 90\%, 95\% and 99\% confidence intervals. } \label{tbl:simulation2}
	\begin{tabular}{|c|c|c|cc|cc|cc|}
		\hline
		\multirow{2}{*}{Code} 
		&\multirow{2}{*}{Method} & \multirow{2}{*}{$L_2$-loss} & 
		\multicolumn{2}{c|}{90\%}
		& \multicolumn{2}{c|}{95\%}
		& \multicolumn{2}{c|}{99\%} \\
		& & & Width & CR & Width & CR & Width & CR\\
	\hline
&  Const   &  0.278 &  0.075 &  0.202 &  0.089 &  0.245 &  0.117 &  0.341 \\
&    &  (0.001) &  (0.001) &  (0.003) &  (0.001) &  (0.004) &  (0.001) &  (0.007) \\
&  Param-Exp   &  0.094 &  0.080 &  0.192 &  0.096 &  0.230 &  0.126 &  0.306 \\
&    &  (0.000) &  (0.001) &  (0.003) &  (0.001) &  (0.003) &  (0.002) &  (0.005) \\
&  Param-Quad   &  0.011 &  0.048 &  0.882 &  0.057 &  0.944 &  0.075 &  0.988 \\
&    &  (0.000) &  (0.001) &  (0.018) &  (0.001) &  (0.012) &  (0.001) &  (0.004) \\
&  RKHS-Cubic   &  0.019 &  0.076 &  0.898 &  0.091 &  0.945 &  0.119 &  0.985 \\
&    &  (0.001) &  (0.001) &  (0.010) &  (0.001) &  (0.007) &  (0.002) &  (0.003) \\
&  Baysian   &  0.025 &  0.116 &  0.948 &  0.138 &  0.963 &  0.182 &  0.978 \\
&    &  (0.001) &  (0.001) &  (0.004) &  (0.001) &  (0.003) &  (0.001) &  (0.002) \\
\hline
&  Const   &  0.279 &  0.072 &  0.196 &  0.086 &  0.240 &  0.113 &  0.328 \\
&    &  (0.001) &  (0.001) &  (0.003) &  (0.001) &  (0.005) &  (0.001) &  (0.006) \\
&  Param-Exp   &  0.095 &  0.079 &  0.186 &  0.094 &  0.223 &  0.124 &  0.297 \\
&    &  (0.000) &  (0.001) &  (0.002) &  (0.001) &  (0.003) &  (0.001) &  (0.004) \\
&  Param-Quad   &  0.011 &  0.048 &  0.889 &  0.058 &  0.939 &  0.076 &  0.982 \\
&    &  (0.001) &  (0.001) &  (0.018) &  (0.001) &  (0.014) &  (0.001) &  (0.008) \\
&  RKHS-Cubic   &  0.019 &  0.074 &  0.891 &  0.088 &  0.940 &  0.115 &  0.986 \\
&    &  (0.001) &  (0.001) &  (0.011) &  (0.001) &  (0.007) &  (0.002) &  (0.003) \\
&  Baysian   &  0.026 &  0.115 &  0.947 &  0.137 &  0.959 &  0.181 &  0.976 \\
&    &  (0.001) &  (0.001) &  (0.004) &  (0.001) &  (0.003) &  (0.001) &  (0.002) \\
\hline
	\end{tabular}
\end{table}

The results are summarized in Tables~\ref{tbl:simulation1} and \ref{tbl:simulation2}. In term of estimation accuracy, the correctly specified parametric calibration model gives the best result, RKHS-Cubic and the Bayesian methods have comparable performance, whereas the Const method performs the worst with no surprise. 
The results for uncertainty quantification are more interesting. In term of the coverage rate close to nominal level, RKHS-Cubic is competitive to the correctly specified parametric calibration in the case of cheap code, and outperforms in the case of expensive code. The confidence intervals produced by mis-specified parametric calibration can have very low coverage. 
Comparing two nonparametric functional calibration methods, the actual confidence interval coverage rate for RKHS-Cubic is closer to the nominal coverage rate, whereas the Bayesian method is conservative and produces much wider intervals with coverage rate higher than the nominal level.

\subsection{Two unidentified models}~\label{sec:unidentified-model}
We consider two settings that the optimal calibration function is not uniquely defined. Since parameter/function estimation is meaningless in this situation, we focus on prediction performance of the competing methods.

\begin{itemize}
	\item The third setting. The physical response is $y^p(x) = 1 + x^3 + \sigma e$ for $x\in [1, 2]$, where $\sigma=0.2$ and $e \sim N(0,1)$.  The computer model is $y^s(x, \bm \theta) = y^s(x, \bm \theta) = \theta_1 x + \theta_2 x^2$ with two calibration parameters $\theta_1$ and $\theta_2$. The functional calibration problem is not identifiable. For example, one possible solution is $\theta_1(x) = 1/x$ and  $\theta_2(x) = x$, another possible solution is $\theta_1(x) = x^2$ and $\theta_2(x) = 1/x^2$. Plugging either solution into the computer model gives a model that matches exactly the physical response function.
		\item The forth setting. The physical response is $y^p(x) =x^3+  \sigma e$ for $x\in[1, 2]$, where $\sigma=0.2$ and $e \sim N(0,1)$. The computer model is $y^s(x, \bm \theta) =  \theta_1 x^{\theta_2}$ with two calibration parameters $\theta_1$ and $\theta_2$. The functional calibration problem is non-identifiable because for any $\alpha\in\mathbb{R}$, the pair $\theta_1(x) = x^{\alpha}$ and $\theta_2(x) = 3 - \alpha$ constitutes a possible solution. 
\end{itemize}

For both the  cheap and expensive code cases in these two settings, the
simulation is repeated 100 times. In each replication, $n=50$ sample
points are generated from on the physical response model with the design points $x_i$'s uniformly generated on the domain. 

The target of prediction is the physical response function $y^p(x)$. To measure the qualify of the calibrated computer model $y^s(x,\hat{\bm \theta}(x))$ in predicting $y^p(x)$, we use the $L_2$-loss for prediction, i.e.,
$\{\int_{\mathcal{X}} [y^p(x)  - y^s(x,\hat{\bm \theta}(x)) ]^2\, dx\}^{1/2} $. 
The average width and average coverage rate for predicting the physical response function can be defined similarly as that for estimating the calibration function, used in Tables 1 and 2.
Again, the integrals are approximated by the Riemann sum with 200 equally spaced points on the domain.

The results for several competing methods are summarized in Table~\ref{tbl:simu3} and Table~\ref{tbl:simu4}. The three methods, Const, Param-Exp, Param-Exp, do not suffer from the identifiability problem since these models impose rigid forms on the calibration function. Our algorithm for our RKHS-Cubic method converges to different local solutions with random initial starting values, but the resulting prediction is rather stable. For the third setting, our RKHS-Cubic method gives the most accurate prediction in terms of the $L_2$-loss for prediction, while the Bayesian method completely fails in prediction in the expensive code case, giving a huge $L_2$-loss. For the fourth setting, our RKHS-Cubic method outperforms other competing methods in prediction except the Const method and the laGP method. That the Const model performs the best in this setting is not surprising, because this setup is in favor of the Const model: The constant calibration parameters $\theta_1 = 1$ and $\theta_2 = 3$ give the optimal solution, and the Const method is constrained to only search over a 2-dimensional parameter set rather than an infinite-dimensional function space for the optimal solution. The laGP method is built on the Const model and thus has its advantage. It is interesting to note that the performance of the laGP method in terms of $L_2$-loss deteriorates significantly from the cheap code to the expensive code case.

\begin{figure}[ht]
     \centering
    \includegraphics[width = 0.48\textwidth]{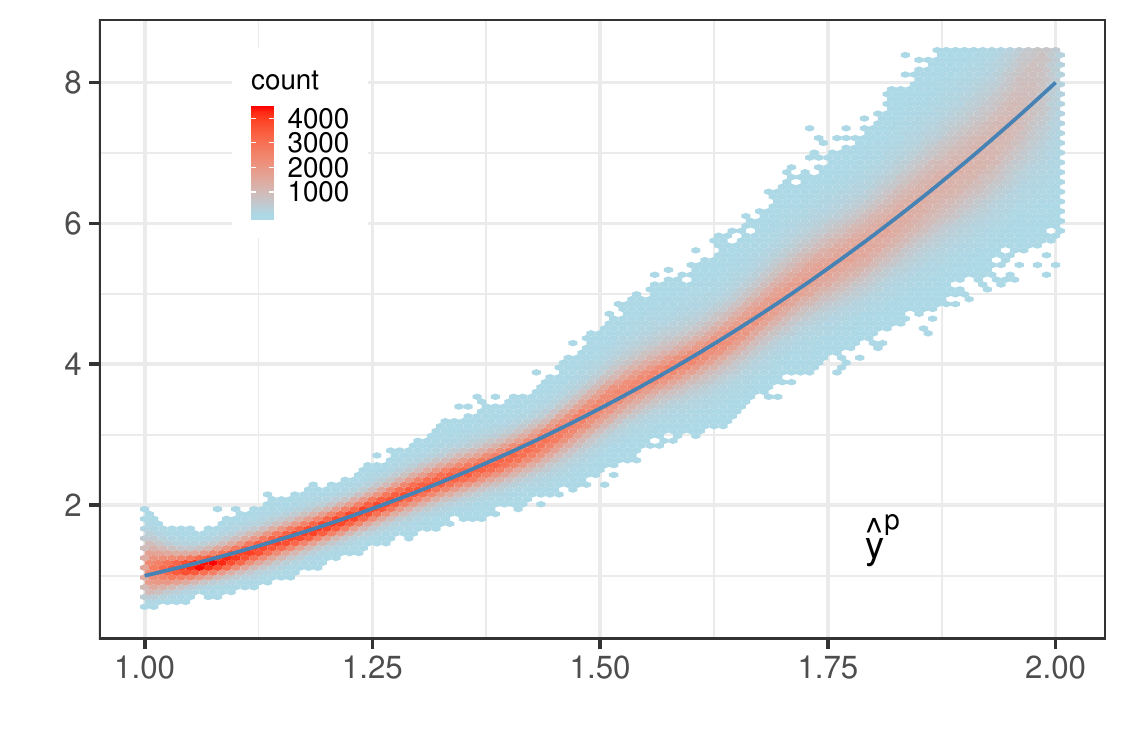}
    \includegraphics[width = 0.48\textwidth]{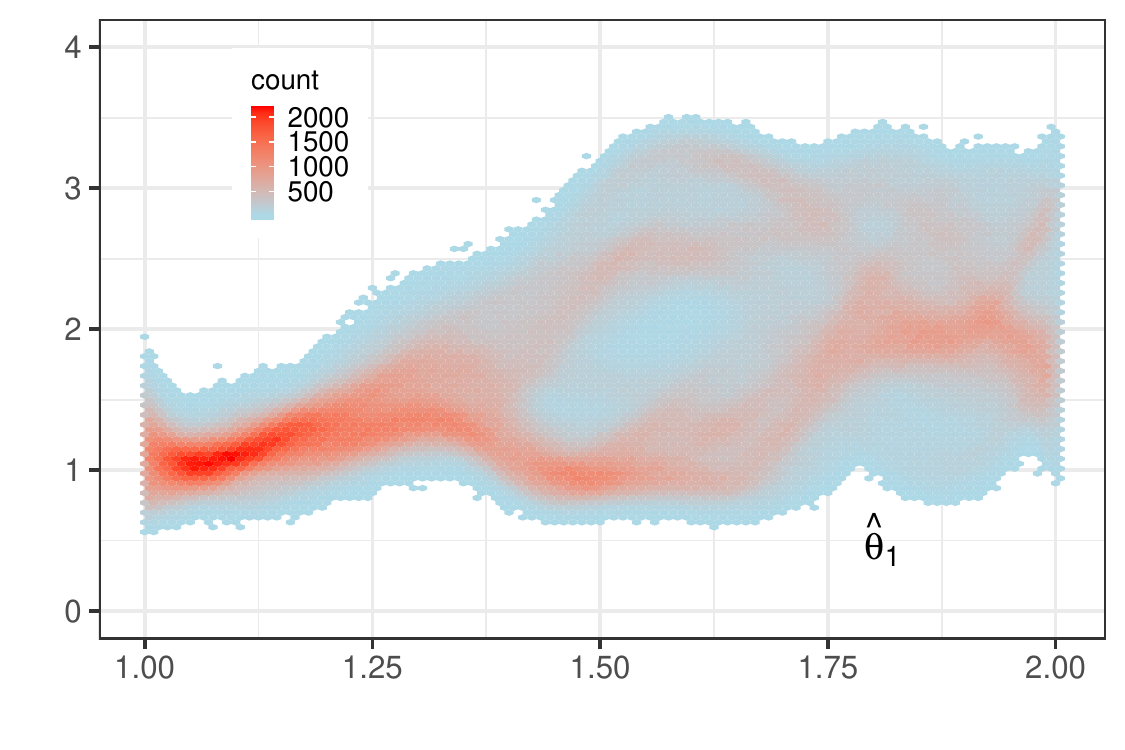}
    \includegraphics[width = 0.48\textwidth]{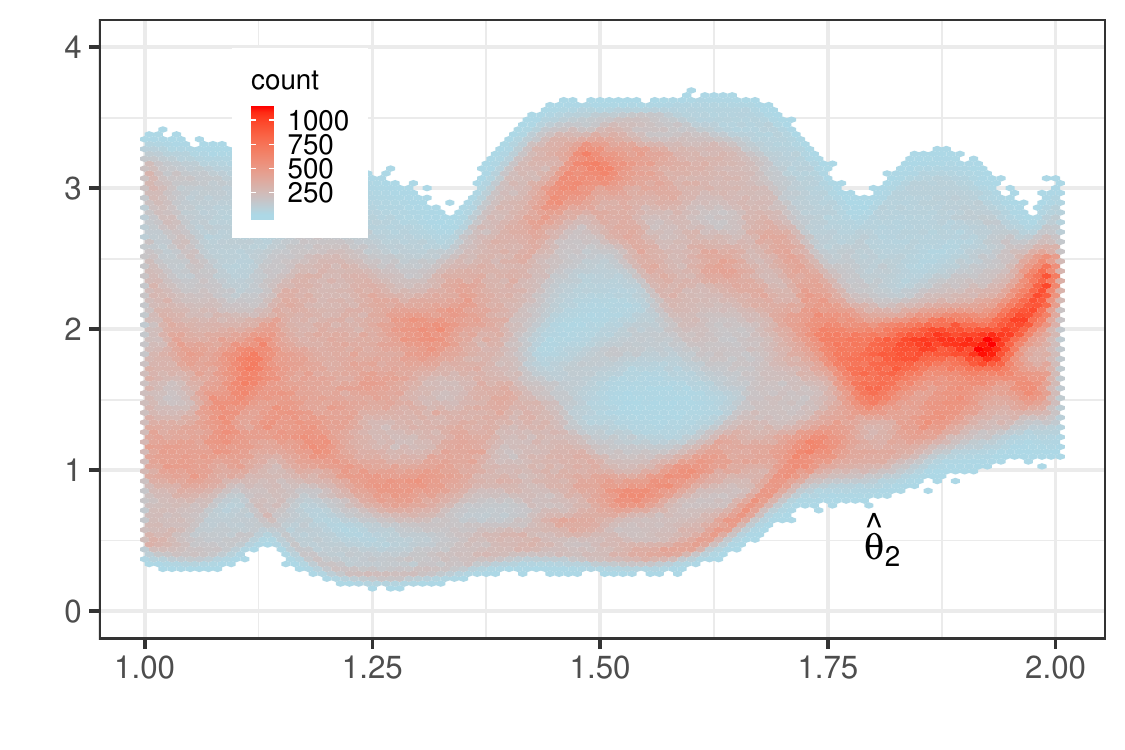}
    \caption{Posterior distributions from the existing Bayesian method for one simulated dataset from Simulation~4. Each panel shows the heat map of the posterior distribution of the underlying function sampled by MCMC after burn-in. 
    The three panels correspond to the physical response $\hat{y}^p$, and two functional calibration parameters $\hat{\theta}_1,\hat{\theta}_2$, respectively. The solid curve in the first panel represents the true physical response curve. }
    \label{fig:reponse:mcmctrace}
\end{figure}

The results for uncertainty quantification in these unidentified settings are more interesting. Our RKHS-Cubic method gives comparatively shorter confidence intervals with actual coverage rate close to the nominal coverage rate (with a slight under-coverage). Even in the cases when the existing Bayesian method \citep{brown2018nonparametric} gives comparable prediction errors, its confidence intervals (for prediction) are substantially wider than those produced by our proposed method. Inspecting the posterior distributions of the functional calibration parameters produced by the MCMC indicates that the MCMC samples diffuse over many local modes; see Figure~\ref{fig:reponse:mcmctrace}. This could be one of the main reasons why the Bayesian method produces substantially wider confidence intervals. Deeper understanding of the behavior of the Bayesian method in unidentified settings needs further research.
The confidence intervals provided by the laGP method are shorter than the corresponding intervals of the Bayesian methods but are still much wider than those provided by our RKHS-Cubic method.

\begin{table}[t]
	\centering
	\caption{Simulation 3. Comparison of methods with the cheap
		code (CC) and expensive code (EC). The mean (and SE) of the $L_2$-loss and of the width and average coverage rate
		(CR) for the level 90\%, 95\% and 99\% confidence intervals. }\label{tbl:simu3}
	\begin{tabular}{|c|c|c|cc|cc|cc|}
		\hline
		\multirow{2}{*}{Code} 
		&\multirow{2}{*}{Method} & \multirow{2}{*}{$L_2$-loss} & 
		\multicolumn{2}{c|}{90\%}
		& \multicolumn{2}{c|}{95\%}
		& \multicolumn{2}{c|}{99\%} \\
		& & & Width & CR & Width & CR & Width & CR\\
		\hline
		CC & Const  &  0.151 &  0.158 &  0.298 &  0.188 &  0.358 &  0.247 &  0.477 \\
		&   &  (0.001) &  (0.002) &  (0.003) &  (0.002) &  (0.004) &  (0.002) &  (0.006) \\
		& Param-Exp   &  0.060 &  0.733 &  0.852 &  0.874 &  0.911 &  1.148 &  0.967 \\
		&   &  (1.016) &  (0.384) &  (0.018) &  (0.457) &  (0.014) &  (0.601) &  (0.009) \\
		& Param-Quad   &  0.052 &  0.181 &  0.899 &  0.216 &  0.944 &  0.284 &  0.985 \\
		&   &  (0.002) &  (0.002) &  (0.015) &  (0.003) &  (0.012) &  (0.003) &  (0.006) \\
		& RKHS-Cubic   &  0.044 &  0.158 &  0.908 &  0.188 &  0.953 &  0.247 &  0.986 \\
		&   &  (0.002) &  (0.002) &  (0.015) &  (0.002) &  (0.011) &  (0.003) &  (0.006) \\
		& Bayesian   &  0.067 &  0.885 &  1.000 &  1.054 &  1.000 &  1.381 &  1.000 \\
		&   &  (0.002) &  (0.001) &  (0.000) &  (0.001) &  (0.000) &  (0.001) &  (0.000) \\
		&  laGP   &  0.057 & 0.667 & 1.000 & 0.795 & 1.000 & 1.044 & 1.000\\
		&   & (0.003) &  (0.007) &  (0.000) &  (0.009) &  (0.000) &  (0.012) &  (0.000)    \\
		\hline
		EC & Const   &  0.216 &  0.125 &  0.184 &  0.149 &  0.223 &  0.196 &  0.300 \\
		&   &  (0.010) &  (0.003) &  (0.006) &  (0.004) &  (0.008) &  (0.005) &  (0.011) \\
		& Param-Exp   &  0.481 &  0.302 &  0.605 &  0.360 &  0.670 &  0.473 &  0.775 \\
		&   &  (0.100) &  (0.040) &  (0.025) &  (0.048) &  (0.025) &  (0.063) &  (0.023) \\
		& Param-Quad   &  0.220 &  0.283 &  0.735 &  0.337 &  0.802 &  0.443 &  0.888 \\
		&   &  (0.025) &  (0.017) &  (0.021) &  (0.021) &  (0.019) &  (0.027) &  (0.014) \\
		& RKHS-Cubic   &  0.114 &  0.311 &  0.879 &  0.370 &  0.915 &  0.486 &  0.954 \\
		&   &  (0.015) &  (0.028) &  (0.017) &  (0.033) &  (0.015) &  (0.043) &  (0.011) \\
		& Bayesian   &  2.326 &  9.000 &  0.891 &  9.982 &  0.916 &  11.206 &  0.934 \\
		&   &  (0.054) &  (0.289) &  (0.016) &  (0.360) &  (0.015) &  (0.470) &  (0.014) \\
		&  laGP   &  0.228 & 0.684 & 0.967 & 0.815 & 0.981 & 1.071 & 0.991\\
		&   &  (0.024) &(0.008) &(0.003) & (0.009) & (0.002) & (0.012) & (0.001) \\
		\hline
	\end{tabular}
\end{table}

\begin{table}[t]
	\centering
	\caption{Original Simulation 4. Comparison of methods with the cheap
		code (CC) and expensive code (EC). The mean (and SE) of the $L_2$-loss and of the width and average coverage rate (CR) for the level 90\%, 95\% and 99\% confidence intervals. }\label{tbl:simu4}
	\begin{tabular}{|c|c|c|cc|cc|cc|}
		\hline
		\multirow{2}{*}{Code} 
		&\multirow{2}{*}{Method} & \multirow{2}{*}{$L_2$-loss} & 
		\multicolumn{2}{c|}{90\%}
		& \multicolumn{2}{c|}{95\%}
		& \multicolumn{2}{c|}{99\%} \\
		& & & Width & CR & Width & CR & Width & CR\\
		\hline
		CC & Const   &  0.035 &  0.125 &  0.883 &  0.149 &  0.928 &  0.196 &  0.982 \\
		&   &  (0.002) &  (0.001) &  (0.023) &  (0.002) &  (0.018) &  (0.002) &  (0.008) \\
		& Param-Exp   &  0.060 &  0.180 &  0.859 &  0.215 &  0.914 &  0.282 &  0.975 \\
		&   &  (0.003) &  (0.004) &  (0.017) &  (0.005) &  (0.013) &  (0.006) &  (0.007) \\
		& Param-Quad   &  0.069 &  0.220 &  0.889 &  0.262 &  0.940 &  0.345 &  0.988 \\
		&   &  (0.002) &  (0.004) &  (0.013) &  (0.005) &  (0.009) &  (0.007) &  (0.004) \\
		& RKHS-Cubic   &  0.058 &  0.181 &  0.879 &  0.216 &  0.928 &  0.284 &  0.977 \\
		&   &  (0.002) &  (0.002) &  (0.017) &  (0.002) &  (0.013) &  (0.003) &  (0.008) \\
		& Bayesian   &  0.076 &  0.820 &  0.998 &  0.978 &  0.999 &  1.286 &  1.000 \\
		&   &  (0.003) &  (0.008) &  (0.001) &  (0.009) &  (0.001) &  (0.013) &  (0.000) \\
		&  laGP   &   0.037 & 0.643 & 1.000 & 0.766 & 1.000 & 1.007 & 1.000  \\
		& &  (0.002) &  (0.007) &  (0.000) &  (0.008) &  (0.000) &  (0.011) &  (0.000) \\
		\hline
		EC& Const   &  0.035 &  0.129 &  0.898 &  0.154 &  0.955 &  0.202 &  0.992 \\
		&   &  (0.002) &  (0.001) &  (0.021) &  (0.002) &  (0.014) &  (0.002) &  (0.004) \\
		& Param-Exp   &  0.082 &  0.175 &  0.740 &  0.208 &  0.805 &  0.274 &  0.886 \\
		&   &  (0.004) &  (0.004) &  (0.020) &  (0.004) &  (0.017) &  (0.006) &  (0.012) \\
		& Param-Quad   &  0.066 &  0.245 &  0.893 &  0.291 &  0.937 &  0.383 &  0.984 \\
		&   &  (0.002) &  (0.014) &  (0.012) &  (0.016) &  (0.009) &  (0.021) &  (0.004) \\
		& RKHS-Cubic   &  0.057 &  0.180 &  0.877 &  0.214 &  0.937 &  0.281 &  0.978 \\
		&   &  (0.002) &  (0.002) &  (0.016) &  (0.003) &  (0.012) &  (0.004) &  (0.006) \\
		& Bayesian   &  0.061 &  0.944 &  0.986 &  1.125 &  0.994 &  1.477 &  0.999 \\
		&   &  (0.002) &  (0.002) &  (0.003) &  (0.002) &  (0.002) &  (0.003) &  (0.000) \\
		&  laGP   & 0.055 & 0.645 & 1.000 & 0.768 & 1.000 & 1.010 & 1.000  \\
		& & (0.002) &  (0.007) &  (0.000) &  (0.009) &  (0.000) &  (0.011) &  (0.000)  \\
		\hline
	\end{tabular}
\end{table}

\section{Real Data: Young's modulus prediction in buckypaper fabrication}~\label{sec_buckypaper}
\citet{Pourhabib2014bp} proposed parametric functional calibration as a method for modulus prediction in buckypaper fabrication. 
Below we apply the proposed nonparametric functional calibration method to the real dataset from \cite{Pourhabib2014bp} and compare with the exponential calibration function used in that work.

Now we provide some background of the problem. Buckypaper is a thin sheet of carbon nanotubes. As far as its mechanical properties are concerned, buckypaper is not directly suitable for most applications. To make it useable for practical purposes, one method is to form composites of buckypaper \citep{tsai2011}, and another approach is to add Poly-vinyl alcohol (PVA) \citep{wang2012b}. In the latter, which yields PVA-treated buckypaper, the goal is to enhance the tensile strength of the final product, measured in terms of Young's modulus.
Practitioners want to understand how the stiffness of the buckypaper, measured in terms of the Young's modulus, is
affected by the addition of PVA in the fabrication process in the presence of other noise variables. A standard approach
is to conduct a set of physical experiments; that is, fabricate a number of buckypapers with varying amounts of the PVA
added, measure the Young's modulus of the resulting buckypaper, and fit a functional relationship between the PVA
input and the stiffness output. Because measuring the Young's modulus requires a process that damages the buckypaper under test, the physical experiments are expensive to conduct, both time-wise and cost-wise. Therefore, a computer model based on a finite element approximation has been developed to numerically calculate the Young's modulus of the buckypaper under a given amount of PVA additive and a few specifications of carbon nanotubes~\citep{wang2012b}. 

\cite{Pourhabib2014bp} reported that this computer model tends to underestimate the
Young's modulus for small amounts of PVA and overestimate the modulus for larger amounts of PVA. Understanding
of the physical process suggests that such a mismatch is caused by the assumption made in the simulation that the \textit{effectiveness} of PVA---i.e., the percentage of the PVA absorbed in the process---stays unchanged as its amount varies. For the computer model outputs to better match the physical experiment outcomes, \cite{Pourhabib2014bp} considered a modified computer model that includes the effectiveness as a calibration parameter. To determine the value of this parameter is
a case of functional calibration because the effectiveness depends on the PVA amount, which is the control variable.

\begin{figure}[t]
	\centering
	\begin{subfigure}[b]{0.45\textwidth}
		\includegraphics[width = \textwidth]{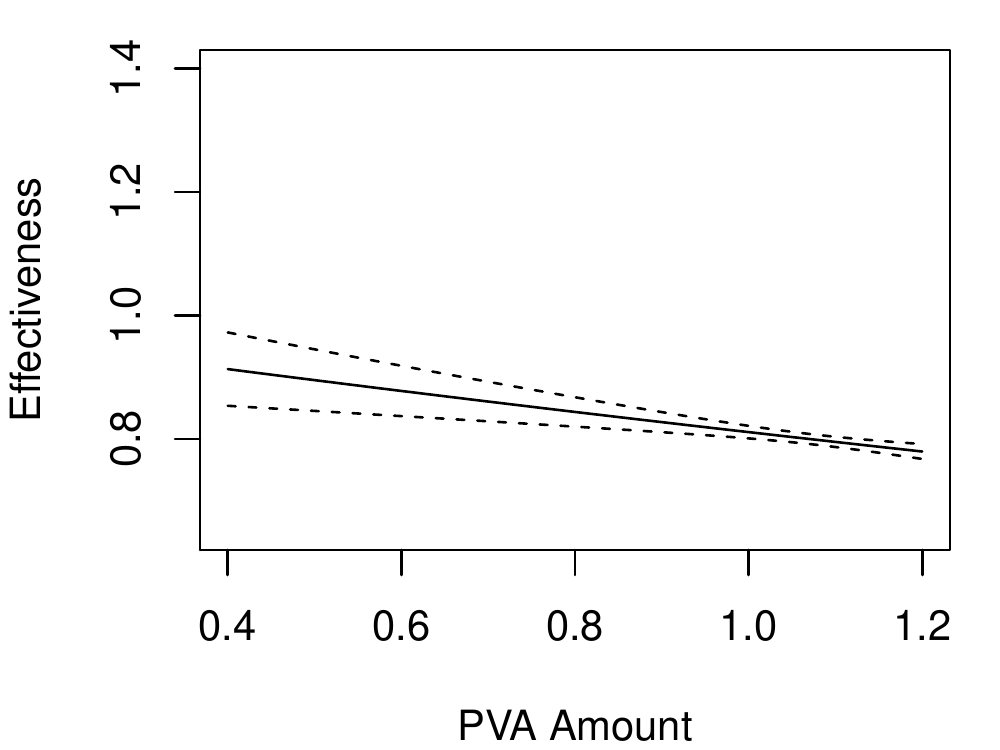}
		\caption{Param-Exp}
		\label{fig:realdata1a}
	\end{subfigure}
	\begin{subfigure}[b]{0.45\textwidth}
		\includegraphics[width = \textwidth]{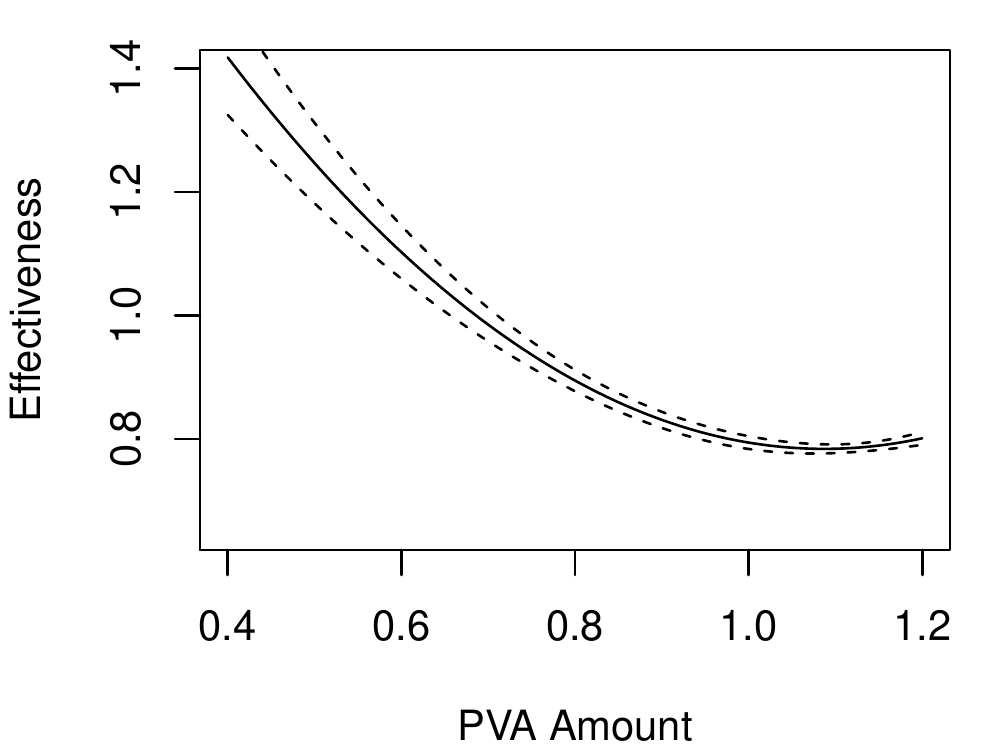}
		\caption{Param-Quad}
		\label{fig:realdata1b}
	\end{subfigure}
	\begin{subfigure}[b]{0.45\textwidth}
		\includegraphics[width = \textwidth]{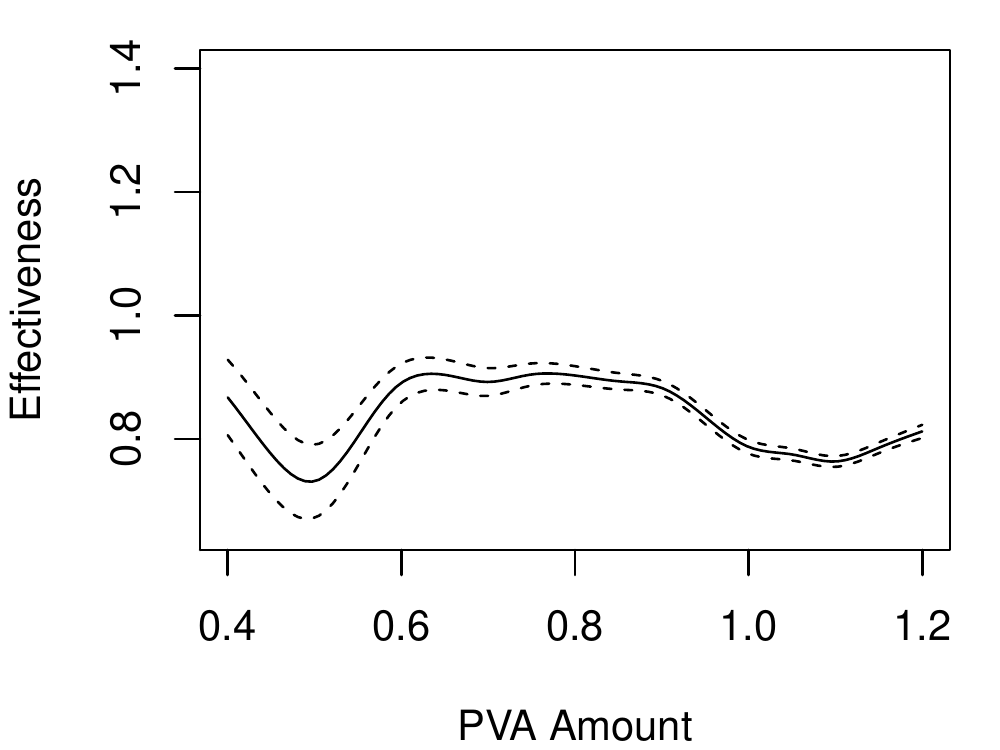}
		\caption{RKHS-Cubic}
		\label{fig:realdata1d}
	\end{subfigure}
	\begin{subfigure}[b]{0.45\textwidth}
		\includegraphics[width = \textwidth]{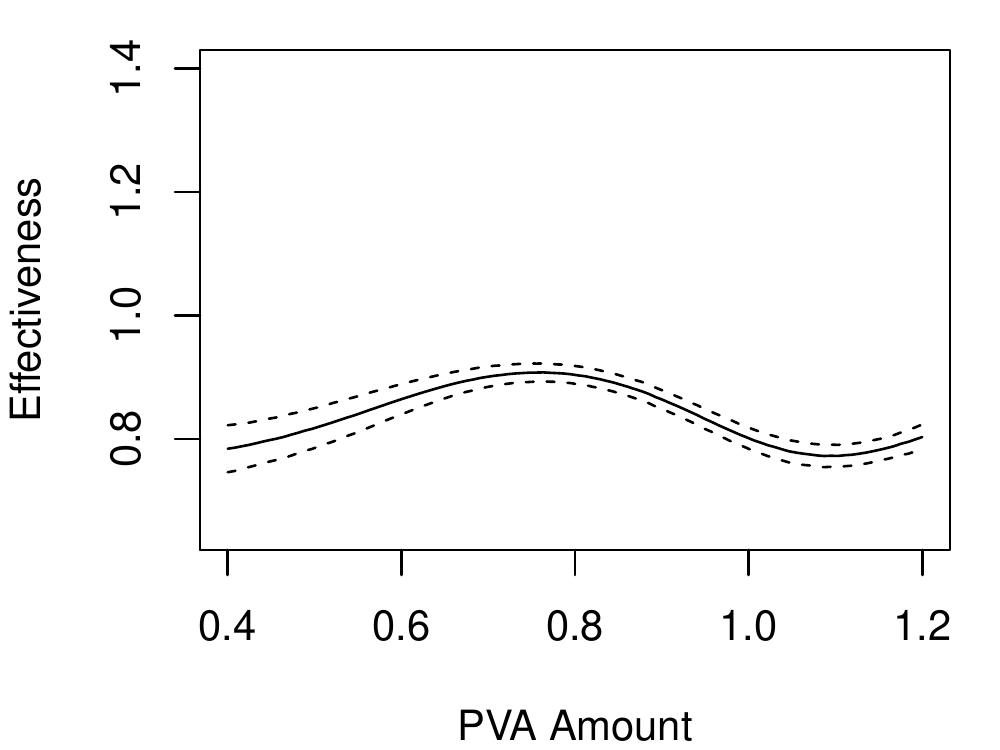}
		\caption{Bayesian}
		\label{fig:realdata1f}
	\end{subfigure}
	\caption{Comparison of four methods for fitting the calibration parameter \textit{effectiveness} as a function of PVA amount. 
		In each panel, the solid line represents the fitted calibration function, and the dashed lines
		represent the uncertainty by one standard deviation from the fitted function. 
		\label{fig:realdata1}} 
\end{figure}

The total number of physical data points is seventeen. To train the emulator model, we used 150 simulated data points from the finite element model. To remove the model bias of the computer model, a constant was subtracted from the physical data such that the physical data and the simulated data have equal average response over their common range of the control variable.  Then the six methods used in the previous section were applied. The fitted calibration functions $\theta(\cdot)$ for various methods are presented in Figure~\ref{fig:realdata1}. 
The Const method and the laGP method have constant calibration parameter and their results are not plotted.
All fitted functions show that the effectiveness is a roughly decreasing function of the PVA amount in the range $[0.7,1.1]$, while the two parametric models indicate steeper decrease. The Param-Quad, RKHS-Cubic and the Bayesian method also show increase of the effectiveness in the PVA range $[1.1,1.2]$. The two functional calibration methods, RKHS-Cubic and the Bayesian method, yield calibration functions with similar general shape, while the result of RKHS-Cubic exhibits greater local variability.

We also used cross-validation to compare the prediction performance of the methods, using the absolute prediction error (APE) as the metric for evaluation. Suppose $y_i^{\mathrm{p}}$ is the observed
response of the $i$-th observation of the physical data, and  $y_i^{\mathrm{cv}}$ is the predicted response using the model fitted by leaving out the $i$-th observation. The corresponding cross-validated APE is computed as  $ \mathrm{APE}_i =  \vert y_i^{\mathrm{p}} - y_i^{\mathrm{cv}}\vert $. From Table~\ref{tbl:realData},
our RKHS-Cubic clearly has the smallest average leave-one-out cross-validated APE. Because we only have seventeen physical data points in total for leave-one-out cross-validation, the standard errors are too big to claim statistical significance. We then performed a leave-two-out cross-validation which randomly leaves two observations out and computed the cross-validated APE for each leave-out observation similar to above. The leave-two-out process was repeated 100 times and the results are presented in Table~\ref{tbl:realData}. The substantially smaller standard errors returned by the leave-two-out cross-validation indicates that the better performance of RHKS-Cubic over competing methods is statistically significant.

\begin{table}[t]
\centering
\caption{Leave-one-out and leave-two-out cross-validation absolute prediction error (APE) for modulus prediction in buckypaper fabrication. The mean and standard error (in parenthesis) of the APEs is presented for each method.}
\label{tbl:realData}
	\begin{tabular}{|c|c|c|c|c|c|c|}
		\hline
		leave-$C$-out&Const   & Param-Exp  & Param-Quad   & 
		RKHS-Cubic  & Bayesian & laGP \\
		\hline
		$C  = 1$ & 130.41& 127.71 & 111.83  & 59.73 & 77.89 & 67.88 \\
		& (19.37) & (25.06) & (17.68)  & (9.33) & (19.21) & (15.13) \\
		\hline
		$C  = 2$ & 129.72& 124.79 & 110.15  & 60.32 & 88.24 & 80.16  \\
		& (5.86) & (7.93) & (4.74)  & (3.09) & (7.61) & (4.72) \\
		\hline
\end{tabular} 
\end{table}

\section{Conclusion}
The existence of different types of dependency among attributes in complex systems is a well-known fact.
In the context of computer experiments, this paper considers the dependency between the calibration parameter and control variables through a functional relationship. In this context, unique challenges are present, such as how to model that dependency, how to obtain predictions, what the theoretical properties of the prediction procedures are, and how to quantify the uncertainty of the predictions. This article develops a frequentist approach to answer such questions.
While existing Bayesian methods require to fully specify a probabilistic model and the computation is based on Markov chain Monte Carlo sampling, our frequentist approach does not rely on a concrete probabilistic model and is based on optimization. We have showed that the frequentist approach
performs competitively in finite sample for both prediction and uncertainty quantification. 

The subject of functional calibration of computer models is clearly widely open for future research. Both frequentist and Bayesian approaches need further development to apply to more sophisticated physical systems. This article provides the first set of asymptotic results on functional calibration of computer models such as consistency and rates of convergence. These results only apply to the frequentist approach. Whether and under what conditions the Bayesian approach has frequentist asymptotic properties are unclear. Asymptotic coverage property of confidence intervals for both frequentist and Bayesian approaches is unstudied and still an important research problem. 

\noindent
{\bf Acknowledgement.} We are grateful to the associate editor and reviewers for many valuable comments which helped significantly improve previous versions of the paper.

\bibliographystyle{chicago}

\newpage
\doublespacing

\vskip 2em
\begin{center}
Supplementary Materials for\\ \vskip 1em
{\Large ``A Reproducing Kernel Hilbert Space Approach to Functional Calibration of Computer Models''}\\
\vskip 1em
{\sc Rui Tuo, Shiyuan He, Arash Pourhabib, Yu Ding and Jianhua Z. Huang}
\end{center}
\vskip 2em

\setcounter{equation}{0}
\setcounter{page}{1}
\setcounter{section}{0}
\renewcommand{\thesection}{S.\arabic{section}}
\renewcommand{\theequation}{S.\arabic{equation}}

This document contains the proofs of the theorems in the paper ``A Reproducing Kernel Hilbert
  Space Approach to Functional Calibration of Computer Models.''
For simplicity in notation, in our proof we only consider the case with a univariate response. The proof can be naturally extended to the general multivariate response case with some complications in notation.

Recall the definition of the estimator of the functional calibration parameter 
\begin{equation}\label{sLC}
  \hat{\theta}_n:=\operatorname*{argmin}\limits_{\theta\in\mathcal{N}^{\Theta}}\frac{1}{n}\sum_{i=1}^n \{y_i^p-\hat{y}_n^s(\bm x_i, \theta(\bm x_i))\}^2+\lambda_n \|\theta\|^2_{\mathcal{N}}.
\end{equation}

\section{Proof of Theorem 1}

	In view of (\ref{sLC}), we have
	\begin{eqnarray}\label{basic_ineq0}
	\begin{split}
	n^{-1}\sum_{i=1}^n \{e_i+\zeta(\bm x_i)-\hat{y}^s_n(\bm x_i,\hat{\theta}_n(\bm x_i))\}^2+\lambda_n \|\hat{\theta}_n\|^2_\mathcal{N}\\
	\leq n^{-1}\sum_{i=1}^n \{e_i+\zeta(\bm x_i)-\hat{y}^s_n(\bm x_i,\theta^*(\bm x_i))\}^2+\lambda_n \|\theta^*\|^2_\mathcal{N}.
	\end{split}
	\end{eqnarray}
	By rearranging (\ref{basic_ineq0}), we obtain
	\begin{eqnarray}\label{basic_ineq1}
	\begin{split}
	&	n^{-1}\sum_{i=1}^n\left\{\left[\zeta(\bm x_i)-\hat{y}^s_n(\bm x_i,\hat{\theta}_n(\bm x_i))\right]^2-\left[\zeta(\bm x_i)-\hat{y}^s_n(\bm x_i,\theta^*(\bm x_i))\right]^2\right\}+\lambda_n\|\hat{\theta}_n\|_{\mathcal{N}}^2
	\\& \leq 2 n^{-1}\sum_{i=1}^n e_i\left\{\hat{y}^s_n(x_i,\hat{\theta}_n(\bm x_i))-\hat{y}^s_n(x_i,\theta^*(\bm x_i))\right\}+\lambda_n\|\theta^*\|_{\mathcal{N}}^2.
	\end{split}
	\end{eqnarray}
	Our derivations in Section~\ref{appendix-sup} shows that
	$$\sup_{f\in\mathcal{N}}\frac{n^{-1}\sum_{i=1}^n e_i f(x_i)}{\|f\|_{\mathcal{N}}}=O_p(n^{-1/2}). $$
	Then by using condition (11) in the statement of the theorem, we have, 
	\begin{eqnarray}
	n^{-1}\sum_{i=1}^n e_i\left\{\hat{y}^s_n(x_i,\hat{\theta}_n(\bm x_i))-\hat{y}^s_n(x_i,\theta^*(\bm x_i))\right\}&=&O_p(n^{-1/2})\|\hat{y}^s_n(x_i,\hat{\theta}_n(\bm x_i))-\hat{y}^s_n(x_i,\theta^*(\bm x_i))\|_{\mathcal{N}}\nonumber\\
	&=&O_p(n^{-1/2})\|\hat{\theta}_n\|_{\mathcal{N}}.\label{innerprod}
	\end{eqnarray}
	
	Using Theorem 2.1 of \cite{van2014uniform} and (11), we find that
	\begin{eqnarray}
		&& n^{-1}\sum_{i=1}^n\left\{\left[\zeta(\bm x_i)-\hat{y}^s_n(\bm x_i,\hat{\theta}_n(\bm x_i))\right]^2-\left[\zeta(\bm x_i)-\hat{y}^s_n(\bm x_i,\theta^*(\bm x_i))\right]^2\right\}\nonumber\\
		&=& \mathrm{PE}(\hat{\bm \theta}_n)- \mathrm{PE}(\bm \theta^*) +O_p(n^{-1/2})(\|\zeta(\bm x)-\hat{y}^s_n(\bm x,\hat{\theta}(\bm x)\|_{\mathcal{N}}^2+\|\zeta(\bm x)-\hat{y}^s_n(\bm x,\theta^*(\bm x)\|_{\mathcal{N}}^2)\label{vandeGeer2014}\\
		&\leq & \mathrm{PE}(\hat{\bm \theta}_n)- \mathrm{PE}(\bm \theta^*) +O_p(n^{-1/2})\|\hat{\theta}_n\|^2_{\mathcal{N}},\nonumber
	\end{eqnarray}
	which, together with (\ref{basic_ineq1}) and (\ref{innerprod}), yields
	\begin{eqnarray}\label{basic_ineq2}
	\mathrm{PE}(\hat{\bm \theta}_n)- \mathrm{PE}(\bm \theta^*)  +(\lambda_n+O_p(n^{-1/2}))\|\hat{\theta}_n\|^2_{\mathcal{N}}\leq O_p(n^{-1/2})\|\hat{\theta}_n\|_{\mathcal{N}}+\lambda_n \|\theta^*\|^2_{\mathcal{N}}.
	\end{eqnarray}
	Because both terms on the left side of (\ref{basic_ineq2}) become positive eventually, the following two inequalities hold for sufficiently large $n$:
	\begin{eqnarray}
	\mathrm{PE}(\hat{\bm \theta}_n)- \mathrm{PE}(\bm \theta^*) \leq O_p(n^{-1/2})\|\hat{\theta}_n\|_{\mathcal{N}}+\lambda_n \|\theta^*\|^2_{\mathcal{N}},\label{inequ1}\\
	(\lambda_n+O_p(n^{-1/2}))\|\hat{\theta}_n\|^2_{\mathcal{N}}\leq O_p(n^{-1/2})\|\hat{\theta}_n\|_{\mathcal{N}}+\lambda_n \|\theta^*\|^2_{\mathcal{N}}.\label{inequ2}
	\end{eqnarray}
	From (\ref{inequ1})-(\ref{inequ2}) and the condition $\lambda_n^{-1}=o_p(n^{1/2})$, we have $\|\hat{\theta}_n\|_{\mathcal{N}}=O_p(1)$ and $\mathrm{PE}(\hat{\bm \theta}_n)- \mathrm{PE}(\bm \theta^*) =O_p(\lambda_n)$. The proof is complete.

\section{Proof of Theorem 2}

The proof is similar to that of Theorem 1. Instead of Using Theorem 2.1 of \cite{van2014uniform}, we now get (\ref{vandeGeer2014}) by applying Condition $1'$.
	
\section{Proof of Theorem 3}\label{proof-theorem1}
The proof proceeds by bounding $\|\theta^*-\hat{\theta}_n\|_n$ and then applying the equivalence of the empirical norm and the $L_2$ norm.
Let
\begin{eqnarray*}
E_n(\theta)&:=&2 n^{-1/2}\sum_{i=1}^n e_i \{\zeta(\bm x_i)-\hat{y}_n^s(\bm x_i,\theta(\bm x_i))\},\\
D_{1 n}(\theta)&:=&n^{-1}\sum_{i=1}^n \{\zeta(\bm x_i)-\hat{y}_n^s(\bm x_i,\theta(\bm x_i))\}^2 ,\\
D_{2 n}(\theta)&:=&\lambda_n \|\theta\|^2_{\mathcal{N}},\\
D_n(\theta)&:=&D_{1 n}(\theta)+D_{2 n}(\theta).
\end{eqnarray*}
The objective function of  the minimization problem (\ref{sLC}) can be written as
\begin{eqnarray*}
&&n^{-1} \sum_{i=1}^n e_i^2 + n^{-1/2} E_n(\theta) + D_n(\theta)\\
& = & n^{-1} \sum_{i=1}^n e_i^2 + n^{-1/2} E_n(\theta) + D_{1n}(\theta) + D_{2n}(\theta)
\end{eqnarray*}
Thus, the solution to  the minimization problem (\ref{sLC}) can also be expressed as
\begin{eqnarray}
\hat{\theta}_n=\operatorname*{argmin}\limits_{\theta\in\mathcal{N}^{\Theta}}\left\{ n^{-1/2}E_n(\theta)+D_n(\theta)\right\}.\label{Estimate}
\end{eqnarray}
Therefore, for all $\delta>0$,
\begin{eqnarray}
\{\|\hat{\theta}_n-\theta^*\|_n>\delta\}\subset& \left\{\inf_{\|\theta-\theta^*\|>\delta}\{n^{-1/2} E_n(\theta)+D_n(\theta)\}<n^{-1/2} E_n(\theta^*)+D_n(\theta^*)\right\},\label{Set}\\
\{\|\hat{\theta}_n\|_{\mathcal{N}}>\delta\}\subset& \left\{\inf_{\|\theta\|_{\mathcal{N}}>\delta}\{n^{-1/2} E_n(\theta)+D_n(\theta)\}<n^{-1/2} E_n(\theta^*)+D_n(\theta^*)\right\}.\label{Set1}
\end{eqnarray}

To bound the probability of the right sides of (\ref{Set}) and (\ref{Set1}), we need to study the property of the two processes indexed by $\theta$, $E_n(\theta)$ and $D_n(\theta)$.
We first look at $E_n(\theta)$.
Let
\[
A(\delta):=\{\theta\in\mathcal{N}^{\Theta}:\|\theta\|_n\leq\delta\},\qquad
B(\rho):=\{\theta\in\mathcal{N}^{\Theta}: \|\theta\|_{\mathcal{N}}\leq \rho\}.
\]
Using results from empirical process theory, we show in Section~\ref{appendix-sup} that
\begin{eqnarray}\label{sup}
E\left[\sup_{\theta\in A(\delta)\cap B(\rho)} |E_n(\theta)-E_n(\theta^*)|\right] \lesssim \delta^{\frac{2\nu-d}{2\nu}}\rho^{\frac{d}{2\nu}},
\end{eqnarray}
where $\lesssim$ denotes that the left side is dominated by the right side up to a universal constant (depending only on $\Omega$ and $\Phi$).
We next consider $D_{1n}(\theta)$, which is a critical component of $D_n(\theta)$.
For notational convenience, we write $\kappa_n=n^{-\frac{2\nu}{2\nu+ d}}$.  In Section~\ref{sect:D_1n}, we show that Conditions~4 and 5 together imply
\begin{eqnarray}\label{eq:D_1n}
D_{1 n}(\theta^*)-D_{1 n}(\theta)
\leq - \omega_0 \, n^{-1}\sum_{i=1}^n\{\theta^*(x_i)-\theta(x_i)\}^2
 + C \kappa_n^{1/2} \|\theta^*-\theta\|_n.
\end{eqnarray}

We now bound the probability of the right sides of (\ref{Set}) and (\ref{Set1}).
To this end, we apply a ``peeling device,'' partitioning the whole space into a countable set of circular rings and applying the maximum inequality on each of them.
Specifically, we partition the set $\{\|\theta-\theta^*\|>\delta\}$ into small pieces.
We now fix $n$.
For $c>0$, define $A_0(c):=[0,c\kappa_n^{1/2}),A_i(c):=[2^{i-1}c\kappa_n^{1/2},2^i c\kappa_n^{1/2}),B_0(c):=[0,c),B_i(c):=[2^{i-1}c,2^i c )$ for $i=1,2,\ldots$. Define
\begin{eqnarray*}
\mathcal{F}_{i j}(c_1,c_2):=\{\theta\in\mathcal{N}^{\Theta}:\|\theta-\theta^*\|_n\in A_i(c_1),\|\theta\|_\mathcal{N}\in B_j(c_2)\}.
\end{eqnarray*}
Then for all $c_1,c_2>0$, $\mathcal{N}^\Theta=\cup_{i,j\geq 0}\mathcal{F}_{i j}(c_1,c_2)$.
Because $\lambda_n\sim \kappa_n$, we can find constant $\xi_1,\xi_2>0$ so that
\begin{eqnarray*}
P(\xi_1\kappa_n\leq\lambda_n\leq \xi_2 \kappa_n)>1-\epsilon/2
\end{eqnarray*}
for all large $n$.
Let $B=\{\xi_1\kappa_n\leq \lambda_n\leq \xi_2 \kappa_n\}$. Then $P(B^c) \leq \epsilon/2$.
Define
\begin{eqnarray*}
P_{i j}(c_1,c_2): & = &P\left(\left\{\inf_{\theta\in\mathcal{F}_{i j}(c_1,c_2)} n^{-1/2}E_n(\theta)+D_n(\theta)<n^{-1/2} E_n(\theta^*)+D_n(\theta^*)\right\} \cap B\right).
\end{eqnarray*}
Then, the probability of the right side of (\ref{Set}) is bounded by $\epsilon/2 + \sum_{i\geq 1,j\geq 0}P_{i j}(c_1,c_2)$ if we let $\delta = c_1 \kappa_n^{1/2}$; and the probability of the right side of (\ref{Set1}) is bounded by $\epsilon/2 + \sum_{i\geq 0,j\geq 1}P_{i j}(c_1,c_2)$ if we let $\delta = c_2$.
Invoking (\ref{Set}) we find that for all $c_1,c_2>0$,
\begin{eqnarray}
P(\|\hat{\theta}_n-\theta^*\|_n>c_1\kappa_n^{1/2})&\leq& \epsilon/2 + \sum_{i\geq 1,j\geq 0}P_{i j}(c_1,c_2),\label{Rate1}\\
P(\|\hat{\theta}_n-\theta^*\|_\mathcal{N}>c_2)&\leq&\epsilon/2 + \sum_{i\geq 0,j\geq 1}P_{i j}(c_1,c_2)\label{Rate2}.
\end{eqnarray}
In Section~\ref{append-prof}, using \eqref{sup} and \eqref{eq:D_1n}, we will show that for any $\epsilon>0$, we can find sufficiently large $c_1,c_2$ so that
\begin{eqnarray}
\sum_{i\geq 1,j\geq 0}P_{i j}(c_1,c_2)\leq \epsilon/2,\qquad \sum_{i\geq 0,j\geq 1}P_{i j}(c_1,c_2)\leq \epsilon/2\label{dominant}
\end{eqnarray}
hold for all large $n$. Combining this with
\eqref{Rate1} and \eqref{Rate2}, we have proved $\|\hat{\theta}_n-\theta^*\|_n=O_p(\kappa_n^{1/2})$ and $\|\hat{\theta}_n\|_\mathcal{N}=O_p(1)$.

The reminder of the proof proceeds by showing that $\|\hat{\theta}_n-\theta^*\|_{L_2(\mathcal{X})}=O_p(n^{-\nu/(2\nu+d)})$.
Lemma 5.16 of \cite{van2000empirical} shows that, if 
\begin{eqnarray}
\sup_{\delta>0} \delta^{\mu}\log N(\delta,\mathcal{A},\|\cdot\|_{L_\infty})<\infty,\label{condition516}
\end{eqnarray}
then  $\|b_n\|_n$ and $\|b_n\|_{L_2(\mathcal{X})}$ are asymptotically equivalent, in the sense that the ratio between the two quantities is bounded away from zero and infinity with probability tending to one as $n$ goes to infinity, for any sequence $\{b_n\}$ on a function space $\mathcal{A}$ with $\|b_n\|_{L_2(\mathcal{X})}\geq C_{\mu,\mathcal{X}} n^{-1/(2+\mu)}$ for some constant $C_{\mu,\mathcal{X}}$ depending only on $\mu$ and $\mathcal{X}$.
By Condition~3, it can be seen that (\ref{condition516}) holds for $\mu=d/\nu$. Note that $O_p(n^{-\nu/(2\nu+d)})= O_p(n^{-1/(2+\mu)})$. Therefore, if $\|\hat{\theta}_n-\theta^*\|_{L_2(\mathcal{X})}< C_{\mu,\mathcal{X}}n^{-\nu/(2\nu+d)}$, there is nothing to prove; otherwise,  $\|\hat{\theta}_n-\theta^*\|_{L_2(\mathcal{X})}$ is upper bounded by a multiple of $\|\hat{\theta}_n-\theta^*\|_n$, then using the proved result
$\|\hat{\theta}_n-\theta^*\|_n=O_p(\kappa_n^{1/2})$,
again we arrive at $\|\hat{\theta}_n-\theta^*\|_{L_2(\mathcal{X})}=O_p(n^{-\nu/(2\nu+d)})$.

\section{Proof of \eqref{sup}}\label{appendix-sup}
Note that
\begin{eqnarray*}
E_n(\theta_1)-E_n(\theta_2)=n^{-1/2}\sum_{i=1}^n 2 e_i \{\hat{y}_n^s(x_i,\theta_2(x_i))-\hat{y}_n^s(x_i,\theta_1(x_i))\}.
\end{eqnarray*}
Because $e_i$ follows a sub-Gaussian distribution with mean 0 and sub-Gaussian parameter $\sigma$, conditional on $X:=(x_1,\ldots,x_n)$, $E_n(\theta_1)-E_n(\theta_2)$ follows a sub-Gaussian distribution with mean zero and sub-Gaussian parameter
\begin{eqnarray}
2 \sigma \left(\frac{1}{n}\sum_{i=1}^n\left\{\hat{y}_n^s(x_i,\theta_2(x_i))-\hat{y}_n^s(x_i,\theta_1(x_i)) \right\}^2\right)^{1/2}.\label{ND}
\end{eqnarray}
Applying mean value theorem to $\hat{y}_n^s$ yields
\[
|\hat{y}_n^s(x_i,\theta_2(x_i))-\hat{y}_n^s(x_i,\theta_1(x_i))|
\leq \|\hat{y}_n^s\|_{C^1(\mathcal{X}\times\Theta)}|\theta_2(x_i)-\theta_1(x_i)|.
\]
Recall that $\kappa_n=n^{-\frac{2\nu}{2\nu+ d}}$.
By the triangle inequality and Condition~4, 
\[
\|\hat{y}_n^s\|_{C^1(\mathcal{X}\times\Theta)}
\leq \kappa_n^{1/2} + \|{y}^s\|_{C^1(\mathcal{X}\times\Theta)}
\leq 2 \|{y}^s\|_{C^1(\mathcal{X}\times\Theta)},
\]
for sufficiently large $n$. Therefore,
\begin{eqnarray}
|\hat{y}_n^s(x_i,\theta_2(x_i))-\hat{y}_n^s(x_i,\theta_1(x_i))|
\leq 2 \|y^s\|_{C^1(\mathcal{X}\times\Theta)}|\theta_2(x_i)-\theta_1(x_i)|.\label{eq:bound1}
\end{eqnarray}
for sufficiently large $n$.
For a space $\mathcal{H}$ with a semi-metric $D$, a stochastic process $X(\cdot)$ over $\mathcal{H}$ is called
sub-Gaussian if $X(s)-X(t)$ is sub-Gaussian with parameter $D(s,t)$
for any $s,t\in\mathcal{H}$. Using \eqref{eq:bound1} to bound \eqref{ND} and noticing
the fact $y^s\in C^1(\mathcal{X}\times\Theta)$, we obtain that,
conditional on $X$, $E_n(\cdot)$ is a sub-Gaussian process with respect to the semi-metric
\begin{eqnarray*}
d_\sigma(\theta_1,\theta_2)=4\sigma \|y^s\|_{C^1(\mathcal{X}\times\Theta)} \|\theta_2 - \theta_1\|_n,
\end{eqnarray*}
for sufficiently large $n$.

Recall the definition of $A(\delta)$ and $B(\rho)$.
Invoking Corollary 2.2.8 of \cite{van1996weak}, we have
\begin{eqnarray*}
E\left[\sup_{\theta\in A(\delta)\cap B(\rho)} |E_n(\theta)-E_n(\theta^*)|\Bigg|X\right]
\lesssim \int_0^{8\delta\sigma\|y^s\|_{C^1(\mathcal{X}\times\Theta)}} \sqrt{\log N(\epsilon,A(\delta)\cap B(\rho),d_\sigma)} d \epsilon.
\end{eqnarray*}
Since
$d_\sigma(\theta_1,\theta_2)\leq 4\sigma \|y^s\|_{C^1(\mathcal{X}\times\Theta)}\|\theta_1-\theta_2\|_{L_\infty(\Omega)},$
the above right-hand side is bounded above by
\begin{eqnarray*}
\int_0^{8\delta\sigma\|y^s\|_{C^1(\mathcal{X}\times\Theta)}} \sqrt{\log N\left(\frac{\epsilon}{4\sigma\|y^s\|_{C^1(\mathcal{X}\times\Theta)}}, B(\rho),L_\infty(\mathcal{X})\right)} d \epsilon.
\end{eqnarray*}
Therefore, using Condition~3 we arrive at,
\begin{eqnarray*}
E\left[\sup_{\theta\in A(\delta)\cap B(\rho)} |E_n(\theta)-E_n(\theta^*)|\Bigg|X\right] \lesssim \delta^{\frac{2\nu-d}{2\nu}}\rho^{\frac{d}{2\nu}}.
\end{eqnarray*}
Taking expectation with respect to the design points yields \eqref{sup}.

\section{Proof of \eqref{eq:D_1n}}\label{sect:D_1n}
Noting the fact that
\begin{eqnarray*}
&&\|f g\|_{C^1}=\|f g\|_{L_\infty}+\|f\nabla g+g\nabla f\|_{L_\infty}\\
&\leq&\|f\|_{L_\infty}\|g\|_{L_\infty}+\|f\|_{L_\infty}\|\nabla g\|_{L_\infty}+\|\nabla f\|_{L_\infty}\|g\|_{L_\infty}\\
&\leq&(\|f\|_{L_\infty}+\|\nabla f\|_{L_\infty})(\|g\|_{L_\infty}+\|\nabla g\|_{L_\infty})=\|f\|_{C^1}\|g\|_{C^1},
\end{eqnarray*}
we obtain together with Condition 4 that
\begin{eqnarray*}
\|(\zeta-y^s)^2-(\zeta-\hat{y}_n^s)^2\|_{C^1(\mathcal{X}\times\Theta)}
&=&\|(2\zeta-y^s-\hat{y}^s_n)(\hat{y}^s_n-y^s)\|_{C^1(\mathcal{X}\times\Theta)}\\
&\leq&\|2\zeta-y^s-\hat{y}^s_n\|_{C^1(\mathcal{X}\times\Theta)}\|\hat{y}^s_n-y^s\|_{C^1(\mathcal{X}\times\Theta)}\\
&\leq&(\|2\zeta-2y^s\|_{C^1(\mathcal{X}\times\Theta)}+\|\hat{y}^s_n-y^s\|_{C^1(\mathcal{X}\times\Theta)}) \|\hat{y}^s_n-y^s\|_{C^1(\mathcal{X}\times\Theta)}\\
&=&(O(1)+O(\kappa_n^{1/2}))O(\kappa_n^{1/2})=O(\kappa_n^{1/2}),
\end{eqnarray*}
i.e., there exists constant $C>0$ so that
\begin{eqnarray}
\|(\zeta-y^s)^2-(\zeta-\hat{y}_n^s)^2\|_{C^1(\mathcal{X}\times\Theta)}\leq C \kappa_n^{1/2}.
\label{EmuCov1}
\end{eqnarray}
Let
\begin{eqnarray*}
D_{1 n}^*(\theta):=n^{-1}\sum_{i=1}^n(\zeta(x_i)-y^s(x_i,\theta(x_i)))^2.
\end{eqnarray*}
For any $\theta\in\mathcal{N}^\Theta$, noting that
\begin{eqnarray*}
D_{1 n}^*(\theta)-D_{1 n}(\theta)=n^{-1}\sum_{i=1}^n(\zeta(x_i)-y^s(x_i,\theta(x_i)))^2-(\zeta(x_i)-\hat{y}^s_n(x_i,\theta(x_i)))^2,
\end{eqnarray*}
we apply mean value theorem to $(\zeta-y^s)^2-(\zeta-\hat{y}^s_n)^2$ and obtain
\begin{equation}\label{increment}
\begin{split}
&|\{D_{1 n}^*(\theta)-D_{1 n}(\theta)\}  - \{D_{1 n}^*(\theta^*)-D_{1 n}(\theta^*)\}|\\
\leq & \|(\zeta-y^s)^2-(\zeta-\hat{y}^s_n)^2\|_{C^1(\mathcal{X}\times\Theta)}
\frac{1}{n} \sum_{i=1}^n  |\theta^*(x_i)-\theta(x_i)|\\
\leq & C\kappa_n^{1/2} \|\theta^*-\theta\|_n,
\end{split}
\end{equation}
where the last inequality follows from (\ref{EmuCov1}) and Cauchy-Schwarz inequality.
Obviously,
\begin{eqnarray*}
&& D_{1 n}(\theta^*)-D_{1 n}(\theta)\\
&\leq& D_{1 n}^*(\theta^*)-D_{1 n}^*(\theta)
+|\{D_{1 n}^*(\theta)-D_{1 n}(\theta)\} - \{D_{1 n}^*(\theta^*)-D_{1 n}(\theta^*)\}|.
\end{eqnarray*}
Using Condition~5 to bound the first term and using (\ref{increment}) to bound the second term of the right-hand side of the above inequality, we obtain
\begin{eqnarray}\label{eq:diff-in-D_1n}
D_{1 n}(\theta^*)-D_{1 n}(\theta)
\leq - \omega_0 \, n^{-1}\sum_{i=1}^n\{\theta^*(x_i)-\theta(x_i)\}^2
 + C \kappa_n^{1/2} \|\theta^*-\theta\|_n.
\end{eqnarray}

\section{Proof of \eqref{dominant}}\label{append-prof}
Note that $P_{i j}(c_1,c_2)$ is upper bounded by
\begin{eqnarray}
p_{ij}& := & P\left(\left\{\inf_{\theta\in\mathcal{F}_{ij}(c_1,c_2)}n^{-1/2}
E_n(\theta)-n^{-1/2}E_n(\theta^*)<D_n(\theta^*)-\inf_{\theta\in\mathcal{F}_{i j}(c_1,c_2)}D_n(\theta)\right\}\cap B \right),\label{Pijepsilon}
\end{eqnarray}
we consider three cases separately: (1) $i,j\geq 1$; (2) $i\geq 1, j=0$; and (3) $i=0, j\geq 1$.
Note that, for all $i,j\geq 0$,
\begin{align}
&D_n(\theta^*)-\inf_{\theta\in\mathcal{F}_{i j}(c_1,c_2)}D_n(\theta)\nonumber\\
\leq& \sup_{\theta\in\mathcal{F}_{i j}(c_1,c_2)}\left\{D_{1 n}(\theta^*)-D_{1 n}(\theta)\right\}
+\{D_{2 n}(\theta^*)- \inf_{\theta\in\mathcal{F}_{i j}(c_1,c_2)}D_{2 n}(\theta)\}\nonumber\\
\begin{split}
\leq&  -  \omega_0\, \inf_{\theta\in\mathcal{F}_{i j}(c_1,c_2)}\frac{1}{n}\sum_{k=1}^n(\theta^*(x_k)-\theta(x_k))^2
+ C \kappa_n^{1/2} \sup_{\theta\in\mathcal{F}_{i j}(c_1,c_2)} \|\theta^*-\theta\|_n\\
& \qquad \qquad \qquad +\{D_{2 n}(\theta^*)- \inf_{\theta\in\mathcal{F}_{i j}(c_1,c_2)}D_{2 n}(\theta)\},\label{rightside}
\end{split}
\end{align}
where the second inequality follows from \eqref{eq:D_1n} or (\ref{eq:diff-in-D_1n}). Using the definition of $\mathcal{F}_{i j}$, we obtain inequalities
\begin{eqnarray}
-\inf_{\theta\in\mathcal{F}_{i j}(c_1,c_2)}\frac{1}{n}\sum_{k=1}^n(\theta^*(x_k)-\theta(x_k))^2\leq\left\{
        \begin{array}{ll}
        -4^{i-1}c_1^2\kappa_n & \hbox{\text{for} $i\geq 1,j\geq 0$,} \\
        0 & \hbox{\text{for} $i= 0,j\geq 1$,}
        \end{array}
        \right.\label{infb1}
\end{eqnarray}
and
\begin{eqnarray}
- \inf\limits_{\theta\in\mathcal{F}_{i j}(c_1,c_2)}D_{2 n}(\theta)\leq\left\{
        \begin{array}{ll}
        -4^{j-1}c_2^2\lambda_n, & \hbox{\text{for} $i\geq 0,j\geq 1$,} \\
        0 & \hbox{\text{for} $i\geq 1,j=0$,}
        \end{array}
        \right.\label{infb2}
\end{eqnarray}
for $i,j\geq 1$.
Let $\rho_0=\|\theta^*\|_\mathcal{N}$. Using (\ref{infb1}) and (\ref{infb2}), (\ref{rightside}) is bounded above by
\begin{eqnarray}
A_{i,j,c_1,c_2,n} :=\left\{
                                   \begin{array}{ll}
                                     \kappa_n\{-\omega_0 4^{i-1}c_1^2+ C2^ic_1\}+\lambda_n\{\rho_0^2- 4^{j-1}c_2^2\}, & \hbox{\text{for} $i,j\geq 1$,} \\
                                     \kappa_n\{-\omega_0 4^{i-1}c_1^2+ C2^ic_1\} +\lambda_n\rho_0^2, & \hbox{\text{for} $i\geq 1,j=0$,} \\
                                     \kappa_n \{Cc_1\}+\lambda_n\{\rho_0^2-4^{ j-1}c_2^2\}, & \hbox{\text{for} $i=0,j\geq 1$,}
                                   \end{array}
                                 \right.
\end{eqnarray}
Then use the condition that $\xi_1\kappa_n\leq \lambda_n\leq \xi_2 \kappa_n$ on set $B$, we find the bound
\begin{eqnarray}
A_{i,j,c_1,c_2,n} \leq \left\{
                                   \begin{array}{ll}
                                     \kappa_n\{-\omega_0 4^{i-1}c_1^2+C2^ic_1+\xi_2\rho_0^2- \xi_1 4^{j-1}c_2^2\}, & \hbox{\text{for} $i,j\geq 1$,} \\
                                     \kappa_n\{-\omega_0 4^{i-1}c_1^2+C2^ic_1 +\xi_2\rho_0^2\}, & \hbox{\text{for} $i\geq 1,j=0$,} \\
                                     \kappa_n \{Cc_1+\xi_2\rho_0^2-\xi_1 4^{ j-1}c_2^2\}, & \hbox{\text{for} $i=0,j\geq 1$,}
                                   \end{array}
                                 \right.
\end{eqnarray}
Clearly, we can choose $c_1,c_2$ sufficiently large so that, for $i,j \geq 1$,
\begin{eqnarray}
-\omega_0 4^{i-1} c_1^2+ C2^i c_1 +\xi_2\rho_0^2\leq -\omega_0 4^{i-1} c_1^2/2,
& \xi_2\rho_0^2-\xi_1 4^{j-1}c_2^2\leq -\xi_1 4^{j-1} c_2^2/2.\label{conditionc1c2}
\end{eqnarray}
We also choose $c_1, c_2$ to satisfy $\xi_1c_2^2/4 > Cc_1$.
In this case, we have
\begin{eqnarray}\label{Aij}
|A_{i,j,c_1,c_2,n}|= - A_{i,j,c_1,c_2,n} \geq \left\{
                                   \begin{array}{ll}
                                     \kappa_n\left\{\omega_0 4^{i-1}c_1^2+ \xi_1 4^{j-1}c_2^2\right\}/2, & \hbox{\text{for} $i,j\geq 1$,} \\
                                     \kappa_n\omega_0 4^{i-1}c_1^2/2, & \hbox{\text{for} $i\geq 1,j=0$,} \\
                                     \kappa_n\xi_1 4^{ j-1}c_2^2/2-\kappa_n Cc_1\geq \kappa_n\xi_1 4^{ j-1}c_2^2/4, & \hbox{\text{for} $i=0,j\geq 1$.}
                                   \end{array}
                                 \right.
\end{eqnarray}
It is easily seen that the choice $c_1=c_2=:c$ satisfies the above conditions provided that $c$ is sufficiently large.
Note that
\begin{eqnarray}
p_{i j}&:=&P\left(\left\{\inf_{\theta\in\mathcal{F}_{i j}(c_1,c_2)}n^{-1/2}E_n(\theta)-n^{-1/2}E_n(\theta^*)<D_n(\theta^*)-\inf_{\theta\in\mathcal{F}_{i j}(c_1,c_2)}D_n(\theta)\right\}\cap B \right)\nonumber\\
&\leq& P\left(\left\{\sup_{\theta\in\mathcal{F}_{i j}(c_1,c_2)}|E_n(\theta)-E_n(\theta^*)|>\sqrt{n}|A_{i,j,c_1,c_2,n}|\right\}\cap B\right)\nonumber\\
&\leq& E\left(\sup_{\theta\in\mathcal{F}_{i j}(c_1,c_2)}|E_n(\theta)-E_n(\theta^*)|\right)\bigg/(\sqrt{n}|A_{i,j,c_1,c_2,n}|)
\end{eqnarray}
We now invoke (\ref{sup}) to find that for all sufficiently large $c$,
\begin{eqnarray}
p_{i,j}&\lesssim&
\left\{
  \begin{array}{ll}
    \frac{(2^i c_1 \kappa^{1/2}_n)^{\frac{2\nu-d}{2\nu}}(2^j c_2)^{\frac{d}{2\nu}}}{\sqrt{n}\kappa_n\left\{\omega_0 4^{i-1}c_1^2+ \xi_1 4^{j-1}c_2^2\right\}/2}, & \hbox{\text{for} $i,j\geq 1$,}  \\
    \frac{(2^ic_1 \kappa^{1/2}_n)^{\frac{2\nu-d}{2\nu}}c_2^{\frac{d}{2\nu}}}{\sqrt{n}\kappa_n\omega_0 4^{i-1}c_1^2/2}, & \hbox{\text{for} $i\geq 1,j=0$,} \\
    \frac{(c_1 \kappa^{1/2}_n)^{\frac{2\nu-d}{2\nu}}(2^jc_2)^{\frac{d}{2\nu}}}{\sqrt{n}\kappa_n \xi_1 4^{j-1}c_2^2/4}, & \hbox{\text{for} $j\geq 1,i=0$,}
  \end{array}
\right.\nonumber\\
&\lesssim&
\begin{cases}
n^{-1/2}c^{-1}\kappa_n^{-\frac{2\nu+d}{4\nu}}\frac{2^{\left\{\frac{2\nu-d}{2\nu}i+\frac{d}{2\nu}j\right\}} }{\omega_0 4^{i}+ \xi_1 4^{j}}, & \hbox{\text{for} $i,j\geq 1$,}  \\
    n^{-1/2}c^{-1}\kappa_n^{-\frac{2\nu+d}{4\nu}}\frac{2^{\frac{2\nu-d}{2\nu}i} }{\omega_0 4^{i}}, & \hbox{\text{for} $i\geq 1,j=0$,} \\
    n^{-1/2}c^{-1}\kappa_n^{-\frac{2\nu+d}{4\nu}}\frac{2^{\frac{d}{2\nu}j} }{ \xi_1 4^{j}}, & \hbox{\text{for} $j\geq 1,i=0$,}
\end{cases}
 \label{Pij}
\end{eqnarray}
where the second inequality follows from the Markov's inequality; the third inequality follows from (\ref{sup}) and (\ref{Aij}).
Because $\kappa_n\sim n^{-\frac{2\nu}{2\nu+d}}$, $n^{-1/2}\kappa_n^{-\frac{2\nu+d}{4\nu}}\sim 1$. Using the inequality
$\omega_0 4^{i}+ \xi_1 4^{j}\geq 2\sqrt{\omega_0\xi_1}2^{i+j},$
we obtain from (\ref{Pij}) (and noticing $\nu > d/2$) that
\begin{eqnarray}\label{Pijbound}
\sum_{i,j\geq 1}p_{i j}&\lesssim & \frac{1}{2} c^{-1} (\omega_0\xi_1)^{-1/2}  \sum_{i,j\geq 1} 2^{-\frac{d}{2\nu}i}2^{-\frac{2\nu-d}{2\nu}j}\nonumber\\
&=&\frac{(\omega_0\xi_1)^{-1/2}(2 c)^{-1}}{2 (1-2^{-\frac{d}{2\nu}})(1-2^{-\frac{2\nu-d}{2\nu}})}=:Q_1(c)<+\infty.
\end{eqnarray}
Elementary calculations also show
\begin{eqnarray}
\sum_{i\geq 1}p_{i 0}&\lesssim & \omega_0^{-1} \sum_{i\geq 1} 2^{-\frac{2\nu+d}{2\nu}i}c^{-1} =\frac{\omega^{-1}_0 2^{-\frac{2\nu+d}{2\nu}}c^{-1} }{1-2^{-\frac{2\nu+d}{2\nu}}}=:Q_2(c)<+\infty,\label{Pi0bound}\\
\sum_{j\geq 1}p_{0 j}&\lesssim &\xi_1^{-1}\sum_{j\geq 1}2^{-\frac{4\nu-d}{2\nu}j}c^{-1} =\frac{\xi_1^{-1}2^{-\frac{4\nu-d}{2\nu}}c^{-1}}{1-2^{-\frac{4\nu-d}{2\nu}}}=:Q_3(c)<+\infty.\label{P0jbound}
\end{eqnarray}
Clearly, one can find sufficiently large $c$ so that
\begin{eqnarray*}
Q_1(c)+Q_2(c)\leq \epsilon/2, \; Q_1(c)+Q_3(c)\leq \epsilon/2,
\end{eqnarray*}
which, together with (\ref{Pijepsilon}), yields (\ref{dominant}).

\section{Proof of Theorem 4}\label{proof-theorem2}
In the proof of Theorem~3 we have actually proved that $\|\hat{\theta}_n-\theta^*\|_n=O_p(n^{-\frac{\nu}{2\nu+d}})$ and $\|\hat{\theta}_n\|_{\mathcal{N}}=O_p(1)$. The proof for these results are also valid for deterministic design points. 
Now it follows from Condition $1'$ that
\[
\|\hat{\theta}_n-\theta^*\|_{L_2(\mathcal{X})}
\lesssim \|\hat{\theta}_n-\theta^*\|_n+n^{-\frac{\nu}{2\nu+d}}\|\hat{\theta}_n-\theta^*\|_{\mathcal{N}}=O_p(n^{-\frac{\nu}{2\nu+d}}),
\]
which completes the proof.


\end{document}